%% file: msr2027.tex
\pdfoutput=1
\documentclass[10pt,conference]{IEEEtran}

\input{preamble}

\input{numbers.tex}
\begin{document}

\title{Stuck on Red?\\Understanding GitHub Actions Failing Streaks}

\author{
	\IEEEauthorblockN{
		Aref Talebzadeh Bardsiri\IEEEauthorrefmark{1},
		Tom Mens\IEEEauthorrefmark{1},
		Alexandre Decan\IEEEauthorrefmark{1},
		Thomas Degueule\IEEEauthorrefmark{2},
		Jean-Rémy Falleri\IEEEauthorrefmark{2}
	}
	\IEEEauthorblockA{
		\IEEEauthorrefmark{1}\textit{Software Engineering Lab,
			University of Mons}, Mons, Belgium\\
		\href{mailto:aref.talebzadehbardsiri@umons.ac.be}
		{aref.talebzadehbardsiri@umons.ac.be},
		\href{mailto:tom.mens@umons.ac.be}
		{tom.mens@umons.ac.be},
		\href{mailto:alexandre.decan@umons.ac.be}
		{alexandre.decan@umons.ac.be}
	}
	\IEEEauthorblockA{
		\IEEEauthorrefmark{2}\textit{Univ. Bordeaux, Bordeaux INP,
			CNRS, LaBRI, UMR 5800}, Bordeaux, France\\
		\href{mailto:thomas.degueule@labri.fr}
		{thomas.degueule@labri.fr},
		\href{mailto:falleri@labri.fr}
		{falleri@labri.fr}
	}
}

\maketitle
\begin{abstract}
GitHub Actions is the {\em de facto} CI/CD service for GitHub repositories. Its workflows should not fail when they are not expected to, nor keep on failing too long. We empirically study \emph{failing streaks}, \ie uninterrupted series of failing runs of a workflow, in \nRunsInScopeShort{} runs of \nWorkflowsInScopeShort{} workflows of \nReposInScopeShort{} GitHub repositories over \windowDays{} days. We study how failing streaks begin and end, and what happens during them, based on the trigger and the commit of each run. We also classify all workflows by their purposes, such as building, testing or deploying, and we inspect the logs of a small sample of failing streaks. Failing runs mostly come in streaks: three quarters of them follow another failing run, and half of the workflows with at least three runs have at least one failing streak. Workflows serving multiple purposes have failing streaks more often, even at the same size. Most failing streaks end within hours, but the one in five that last more than a day account for the lion's share of the time that workflows spend failing. A third of the failing streaks start, and two in five end, without any change to the repository. In a sample of 100 failing streaks, those that end this way most often fail because of network issues or an external service. These findings suggest that maintainers should not assume that a failure comes from the last commit. GitHub and tool builders could flag failures that occur without any change to the repository, and alert the active maintainers when a workflow stays failing for more than a day.
\end{abstract}

\input{Input/introduction.tex}

\input{Input/related.tex}

\input{Input/methodology.tex}
\input{Input/rq1.tex}
\input{Input/rq2.tex}

\input{Input/rq3.tex}

\input{Input/rq4.tex}
\input{Input/discussion.tex}

\input{Input/threats.tex}

\input{Input/conclusion.tex}

\input{Input/data_availability.tex}

\newpage
\balance
\bibliographystyle{IEEEtran}
\bibliography{biblio}
\end{document}

%% file: preamble.tex
\usepackage{xspace}
\usepackage{mdframed}
\usepackage{xurl}
\usepackage[disable]{todonotes}
\usepackage{listings}
\usepackage[inline]{enumitem}
\usepackage{graphicx}
\usepackage{xcolor}
\usepackage{booktabs}
\usepackage{microtype}
\usepackage[hidelinks]{hyperref}
\usepackage{balance}
\usepackage{amsmath,amssymb}
\usepackage{cite}
\usepackage[scaled=.8]{beramono}
\usepackage{pifont}%
\makeatletter

\definecolor{ymlkey}{HTML}{1B2733}
\definecolor{ymlval}{HTML}{0B6E5B}
\definecolor{ymlpunct}{HTML}{9AA1A9}
\definecolor{ymlcomment}{HTML}{7A828B}
\definecolor{ymlrule}{HTML}{3F4A56}
\definecolor{ymlnum}{HTML}{A8AEB5}
\definecolor{callout}{RGB}{0,120,120}

\definecolor{mygreen}{HTML}{04481b}
\definecolor{myred}{HTML}{910a0a}

\newcommand{\success}{{\color{mygreen}\ding{51}}\xspace}
\newcommand{\failure}{{\color{myred}\ding{55}}\xspace}

\newcommand{\etal}{et al.\xspace}
\newcommand{\ie}{i.e.,\xspace}
\newcommand{\eg}{e.g.,\xspace}
\newcommand{\fig}[1]{Figure~\ref{#1}}
\newcommand{\tab}[1]{Table~\ref{#1}}
\newcommand{\sect}[1]{Section~\ref{#1}}

\newcommand{\cicd}{CI/CD\xspace}
\newcommand{\github}{GitHub\xspace}

\newcommand{\gha}{GHA\xspace}

\newcommand{\ymlKeyStyle}{\ttfamily\color{ymlkey}\bfseries}
\newcommand{\ymlValStyle}{\ttfamily\color{ymlval}\mdseries}
\newcommand{\ymlSepStyle}{\ttfamily\color{ymlpunct}\mdseries}

\newcommand{\language@yaml}{yaml}
\lstdefinelanguage{yaml}{
  sensitive=false,
  basicstyle=\ymlKeyStyle,
  keywords={true,false,null},
  keywordstyle=\ymlValStyle\itshape,
  comment=[l]{\#},
  commentstyle=\ttfamily\color{ymlcomment}\itshape,
  morestring=[b]",
  morestring=[b]',
  stringstyle=\ymlValStyle,
  moredelim=**[il][\ymlSepStyle{:}\ymlValStyle]{:},
  moredelim=[l][\ymlValStyle]{\&},
  moredelim=[l][\ymlValStyle]{*},
  literate=
    {-\ }{{\mbox{\ymlSepStyle-}\ }}2
    {---}{{\mbox{\ymlSepStyle---}}}3,
}
\lst@AddToHook{EveryLine}{\ifx\lst@language\language@yaml\ymlKeyStyle\fi}

\lstdefinestyle{yamlstyle}{
  language=yaml,
  basicstyle=\ymlKeyStyle\footnotesize,
  breaklines=false,
  breakatwhitespace=true,
  breakindent=1.2em,
  columns=flexible,
  keepspaces=true,
  showstringspaces=false,
  tabsize=2,
  frame=none,
  numbers=left,
  numberstyle=\ttfamily\tiny\color{ymlnum},
  numbersep=9pt,
  xleftmargin=2.2em,
  aboveskip=3pt,
  belowskip=3pt,
  captionpos=b,
}

\newcommand{\anonymise}[1]{<anonymised>}

\newcommand{\summary}[1]{\begin{mdframed}[backgroundcolor=black!5, topline=false, rightline=false,  leftline=true, bottomline=false,  linewidth=2pt, linecolor=black!70, nobreak=false, innertopmargin=4pt, innerbottommargin=4pt, skipabove=4pt, skipbelow=4pt, innerleftmargin=4pt, innerrightmargin=4pt]\noindent{#1}\end{mdframed}}

\global\mdfdefinestyle{exampledefault}{linewidth=1pt, linecolor=lightgray!80, backgroundcolor=lightgray!20, innerleftmargin=4pt, innerrightmargin=4pt, innertopmargin=3pt, innerbottommargin=3pt, nobreak=true,roundcorner=5pt}
\newenvironment{custombox}{\smallskip\begin{mdframed}[style=exampledefault]}{\end{mdframed}\smallskip}

%% file: numbers.tex
\newcommand{\nAllRuns}{130,338,516}

\newcommand{\nAllWorkflows}{226,797}

\newcommand{\nAllRepositories}{47,695}

\newcommand{\nRunsInScope}{30,923,432}
\newcommand{\nRunsInScopeShort}{30.9M+}
\newcommand{\nWorkflowsInScope}{154,846}
\newcommand{\nWorkflowsInScopeShort}{154K+}
\newcommand{\nReposInScope}{43,936}
\newcommand{\nReposInScopeShort}{43K+}
\newcommand{\windowStart}{2026-02-01}
\newcommand{\windowEnd}{2026-09-26}
\newcommand{\windowDays}{238}
\newcommand{\windowDaysShort}{238}
\newcommand{\nFailingStreaks}{710,654}

\newcommand{\medianFailingStreaksPerWorkflow}{3}

\newcommand{\nFailingStreaksStartedBySchedule}{130,199}

\newcommand{\pctFailingStreaksStartedBySchedule}{18.3}
\newcommand{\nFailingStreaksStartedByPush}{378,701}

\newcommand{\pctFailingStreaksStartedByPush}{53.3}
\newcommand{\nFailingStreaksStartedByWorkflowRun}{100,846}

\newcommand{\pctFailingStreaksStartedByWorkflowRun}{14.2}

\newcommand{\nFailingStreaksStartedByWorkflowDispatch}{64,040}

\newcommand{\pctFailingStreaksStartedByWorkflowDispatch}{9.0}

\newcommand{\nFailingStreaksStartedByOther}{36,868}

\newcommand{\pctFailingStreaksStartedByOther}{5.2}

\newcommand{\nFailingStreaksPerWorkflow}{71,231}

\newcommand{\pctStartedOnDiffCommitEndedOnDiffCommit}{51.4}
\newcommand{\pctStartedOnDiffCommitEndedOnSameCommit}{14.9}
\newcommand{\pctStartedOnSameCommitEndedOnDiffCommit}{6.8}
\newcommand{\pctStartedOnSameCommitEndedOnSameCommit}{27.0}

\newcommand{\pctFailingEndsSameCommit}{41.8}

\newcommand{\pctOneEventFailingStreak}{92.8}

\newcommand{\pctWorkflowsWithFailingRun}{59.2}

\newcommand{\medianRunsPerWorkflow}{38}

\newcommand{\meanRunsPerWorkflow}{199.7}

\newcommand{\medianRunSeconds}{60}

\newcommand{\meanRunSeconds}{832.9}

\newcommand{\pctReposWithFailures}{76.3}

\newcommand{\pctBreakPairsProjectFilesOnly}{82.4}
\newcommand{\pctBreakPairsWorkflowAndProjectFiles}{12.9}

\newcommand{\pctBreakPairsOwnWorkflowFile}{7.2}

\newcommand{\pctBaselinePairsProjectFilesOnly}{87.5}
\newcommand{\pctBaselinePairsWorkflowAndProjectFiles}{7.4}

\newcommand{\pctBaselinePairsOwnWorkflowFile}{2.8}
\newcommand{\pctWorkflowsWithFailingStreak}{51.0}
\newcommand{\pctSameCommitStartPush}{0.2}
\newcommand{\pctSameCommitStartSchedule}{68.7}
\newcommand{\pctSameCommitStartWorkflowRun}{78.6}

\newcommand{\pctSameCommitStartWorkflowDispatch}{67.6}

\newcommand{\pctSameCommitStartOther}{74.0}
\newcommand{\pctSameCommitStartAll}{33.8}
\newcommand{\minRunsFirstFifth}{1}

\newcommand{\maxRunsFirstFifth}{7}

\newcommand{\minRunsSecondFifth}{8}

\newcommand{\maxRunsSecondFifth}{25}

\newcommand{\minRunsThirdFifth}{26}

\newcommand{\maxRunsThirdFifth}{65}

\newcommand{\minRunsFourthFifth}{66}

\newcommand{\maxRunsFourthFifth}{212}

\newcommand{\minRunsLastFifth}{213}

\newcommand{\nWorkflowsFirstFifth}{17,715}

\newcommand{\pctWithFailingStreakFirstFifth}{11.2}
\newcommand{\medianFailingStreaksFirstFifth}{1}

\newcommand{\nWorkflowsSecondFifth}{29,661}

\newcommand{\pctWithFailingStreakSecondFifth}{33.3}
\newcommand{\medianFailingStreaksSecondFifth}{1}

\newcommand{\nWorkflowsThirdFifth}{30,703}

\newcommand{\pctWithFailingStreakThirdFifth}{49.6}
\newcommand{\medianFailingStreaksThirdFifth}{2}

\newcommand{\nWorkflowsFourthFifth}{30,745}

\newcommand{\pctWithFailingStreakFourthFifth}{66.3}
\newcommand{\medianFailingStreaksFourthFifth}{3}

\newcommand{\nWorkflowsLastFifth}{30,878}

\newcommand{\pctWithFailingStreakLastFifth}{76.9}
\newcommand{\medianFailingStreaksLastFifth}{6}

\newcommand{\nWorkflowsAllWorkflows}{139,702}

\newcommand{\pctWithFailingStreakAllWorkflows}{51.0}
\newcommand{\medianFailingStreaksAllWorkflows}{3}

\newcommand{\pctDifferentCommitStartAll}{66.2}

\newcommand{\nStreaksWithPairAll}{240,609}

\newcommand{\pctStreaksNewCommitWhileFailing}{62.7}
\newcommand{\nDuringPairsWithDiff}{592,153}

\newcommand{\pctDuringPairsProjectFilesOnly}{83.7}
\newcommand{\pctDuringPairsWorkflowAndProjectFiles}{10.5}

\newcommand{\pctDuringPairsOwnWorkflowFile}{5.7}
\newcommand{\pctRunsFailing}{10.1}

\newcommand{\pctFailingRunsAfterFailingRun}{75.8}
\newcommand{\medianRunsWorkflowsNeverFail}{17}

\newcommand{\medianRunsWorkflowsWithFailingRun}{70}

\newcommand{\pctRunsInWorkflowsNeverFail}{12.0}

\newcommand{\pctReposWithFailingStreak}{68.0}

\newcommand{\pctFailingStreaksEndedByPush}{55.0}
\newcommand{\pctSameCommitEndPush}{13.5}

\newcommand{\pctSameCommitEndSchedule}{72.6}

\newcommand{\pctSameCommitEndWorkflowRun}{82.0}

\newcommand{\pctSameCommitEndAll}{41.8}

\newcommand{\medianHoursStartedOnSameCommitEndedOnDiffCommit}{8.5}
\newcommand{\pctLongerThanDayStartedOnSameCommitEndedOnDiffCommit}{31.0}
\newcommand{\medianHoursStartedOnSameCommitEndedOnSameCommit}{0.6}

\newcommand{\pctFailingStreaksSameTriggerBothEnds}{93.7}
\newcommand{\nFixPairsWithDiff}{411,596}

\newcommand{\pctFixPairsProjectFilesOnly}{81.2}
\newcommand{\pctFixPairsWorkflowAndProjectFiles}{12.4}

\newcommand{\pctFixPairsOwnWorkflowFile}{9.2}

\newcommand{\meanFailingRunsPerStreakFirstFifth}{1.3}

\newcommand{\pctRunsFailingLastFifth}{9.4}
\newcommand{\meanFailingRunsPerStreakLastFifth}{2.9}
\newcommand{\medianFailingHours}{1.9}
\newcommand{\medianFailingHoursPerWorkflow}{7.7}
\newcommand{\pctStreaksLongerThanDay}{19.5}
\newcommand{\pctFailingTimeLongerThanDay}{91.7}
\newcommand{\pctStreaksLongerThanWeek}{4.5}
\newcommand{\pctFailingTimeLongerThanWeek}{67.7}

\newcommand{\pctNewCommitLongSchedule}{74.4}

\newcommand{\pctScheduleStartEndedByPush}{13.6}

\newcommand{\pctSameCommitPassesAfterOne}{57.7}

\newcommand{\pctSameCommitPassesAfterSixPlus}{3.5}

\newcommand{\pctStartupFailureOfFailingRuns}{2.3}
\newcommand{\nRunsWithHiddenAttempts}{147,638}

\newcommand{\nHiddenAttempts}{204,106}

\newcommand{\nHiddenAttemptsInScope}{182,871}

\newcommand{\maxPathsSmall}{20}

\newcommand{\minPathsMedium}{21}

\newcommand{\maxPathsMedium}{40}

\newcommand{\minPathsLarge}{41}

\newcommand{\nWorkflowsSizeTable}{139,049}

\newcommand{\spearmanLabelsPaths}{0.56}
\newcommand{\medianPathsOneLabel}{18}

\newcommand{\medianPathsFourPlusLabels}{76}

\newcommand{\pctSizeTableMainTriggerPush}{65.3}
\newcommand{\pctSizeTableMainTriggerSchedule}{21.0}
\newcommand{\pctWithStreakSmallOneLabel}{32.4}
\newcommand{\pctWithStreakSmallTwoLabels}{41.2}
\newcommand{\pctWithStreakSmallThreeLabels}{46.4}
\newcommand{\pctWithStreakSmallFourPlusLabels}{54.8}
\newcommand{\pctWithStreakMediumOneLabel}{43.4}
\newcommand{\pctWithStreakMediumTwoLabels}{49.0}
\newcommand{\pctWithStreakMediumThreeLabels}{54.9}
\newcommand{\pctWithStreakMediumFourPlusLabels}{58.7}
\newcommand{\pctWithStreakLargeOneLabel}{59.4}
\newcommand{\pctWithStreakLargeTwoLabels}{61.6}
\newcommand{\pctWithStreakLargeThreeLabels}{68.4}
\newcommand{\pctWithStreakLargeFourPlusLabels}{75.9}
\newcommand{\nWorkflowsPurposesOne}{47,055}

\newcommand{\pctWithStreakPurposesOne}{38.2}
\newcommand{\medianStreaksPurposesOne}{2}

\newcommand{\nWorkflowsPurposesTwo}{46,693}

\newcommand{\pctWithStreakPurposesTwo}{50.0}
\newcommand{\medianStreaksPurposesTwo}{2}

\newcommand{\nWorkflowsPurposesThree}{25,251}

\newcommand{\pctWithStreakPurposesThree}{60.4}
\newcommand{\medianStreaksPurposesThree}{3}

\newcommand{\nWorkflowsPurposesFourPlus}{20,050}

\newcommand{\pctWithStreakPurposesFourPlus}{72.1}
\newcommand{\medianStreaksPurposesFourPlus}{5}

\newcommand{\purposeSetOneA}{pr-and-issues}
\newcommand{\nWorkflowsPurposeSetOneA}{9,145}

\newcommand{\pctWithStreakPurposeSetOneA}{35.0}
\newcommand{\medianStreaksPurposeSetOneA}{2}

\newcommand{\purposeSetOneB}{quality}
\newcommand{\nWorkflowsPurposeSetOneB}{7,602}

\newcommand{\pctWithStreakPurposeSetOneB}{35.9}
\newcommand{\medianStreaksPurposeSetOneB}{2}

\newcommand{\purposeSetOneC}{test}
\newcommand{\nWorkflowsPurposeSetOneC}{6,780}

\newcommand{\pctWithStreakPurposeSetOneC}{48.6}
\newcommand{\medianStreaksPurposeSetOneC}{2}

\newcommand{\purposeSetTwoA}{build+test}
\newcommand{\nWorkflowsPurposeSetTwoA}{9,631}

\newcommand{\pctWithStreakPurposeSetTwoA}{52.8}
\newcommand{\medianStreaksPurposeSetTwoA}{2}

\newcommand{\purposeSetTwoB}{build+release}
\newcommand{\nWorkflowsPurposeSetTwoB}{8,748}

\newcommand{\pctWithStreakPurposeSetTwoB}{50.9}
\newcommand{\medianStreaksPurposeSetTwoB}{2}

\newcommand{\purposeSetTwoC}{report+test}
\newcommand{\nWorkflowsPurposeSetTwoC}{3,221}

\newcommand{\pctWithStreakPurposeSetTwoC}{55.6}
\newcommand{\medianStreaksPurposeSetTwoC}{3}

\newcommand{\purposeSetThreeA}{build+report+test}
\newcommand{\nWorkflowsPurposeSetThreeA}{4,010}

\newcommand{\pctWithStreakPurposeSetThreeA}{63.7}
\newcommand{\medianStreaksPurposeSetThreeA}{4}

\newcommand{\purposeSetThreeB}{build+quality+test}
\newcommand{\nWorkflowsPurposeSetThreeB}{3,876}

\newcommand{\pctWithStreakPurposeSetThreeB}{60.3}
\newcommand{\medianStreaksPurposeSetThreeB}{3}

\newcommand{\purposeSetThreeC}{build+release+test}
\newcommand{\nWorkflowsPurposeSetThreeC}{2,848}

\newcommand{\pctWithStreakPurposeSetThreeC}{61.4}
\newcommand{\medianStreaksPurposeSetThreeC}{3}

\newcommand{\purposeSetFourPlusA}{build+quality+report+test}
\newcommand{\nWorkflowsPurposeSetFourPlusA}{1,979}

\newcommand{\pctWithStreakPurposeSetFourPlusA}{69.0}
\newcommand{\medianStreaksPurposeSetFourPlusA}{4}

\newcommand{\purposeSetFourPlusB}{build+quality+release+test}
\newcommand{\nWorkflowsPurposeSetFourPlusB}{1,196}

\newcommand{\pctWithStreakPurposeSetFourPlusB}{67.3}
\newcommand{\medianStreaksPurposeSetFourPlusB}{3}

\newcommand{\purposeSetFourPlusC}{build+docs+quality+test}
\newcommand{\nWorkflowsPurposeSetFourPlusC}{789}

\newcommand{\pctWithStreakPurposeSetFourPlusC}{66.4}
\newcommand{\medianStreaksPurposeSetFourPlusC}{4}

\newcommand{\pctLengthFourPlus}{12.0}

\newcommand{\pctLengthFourPlusPerWorkflow}{12.3}
\newcommand{\medianHoursOneRun}{1.0}
\newcommand{\medianHoursTwoPlusRuns}{6.8}

\newcommand{\medianHoursOnePurpose}{0.8}
\newcommand{\medianHoursFourPlusPurposes}{2.4}
\newcommand{\medianHoursOnePurposePerWorkflow}{6.5}
\newcommand{\medianHoursFourPlusPurposesPerWorkflow}{7.2}
\newcommand{\nWorkflowsTopOnePctOnePurpose}{180}

\newcommand{\pctStreaksTopOnePctOnePurpose}{44.2}
\newcommand{\medianMinutesTopOnePctOnePurpose}{9}

\newcommand{\nWorkflowsCanHaveStreak}{139,702}

\newcommand{\nWorkflowsOneOrTwoRuns}{15,144}

\newcommand{\pctWorkflowsAlwaysFail}{6.0}
\newcommand{\nLogBuildTestNewNew}{22}

\newcommand{\nLogBuildTestSameNew}{17}

\newcommand{\nLogSameSameOtherPurposes}{16}

\newcommand{\nLogSameSameOtherPurposesScheduleOrWorkflowRun}{13}

\newcommand{\nLogProjectNewNew}{14}

\newcommand{\nLogWorkflowNewNew}{8}

\newcommand{\nLogOtherNewNew}{3}

\newcommand{\nLogProjectNewSame}{10}

\newcommand{\nLogWorkflowNewSame}{13}

\newcommand{\nLogOtherNewSame}{2}

\newcommand{\nLogProjectSameNew}{12}

\newcommand{\nLogWorkflowSameNew}{8}

\newcommand{\nLogOtherSameNew}{5}

\newcommand{\nLogProjectSameSame}{6}

\newcommand{\nLogWorkflowSameSame}{16}

\newcommand{\nLogOtherSameSame}{3}

\newcommand{\nLogProjectAll}{42}

\newcommand{\nLogWorkflowAll}{45}

\newcommand{\nLogOtherAll}{13}

\newcommand{\nLogStreaks}{100}

\newcommand{\nLogTestsFailNewNew}{7}

\newcommand{\nLogRestProjectNewNew}{7}

\newcommand{\nLogWorkflowConfigurationNewNew}{1}

\newcommand{\nLogNetworkOrServiceNewNew}{3}

\newcommand{\nLogRestWorkflowNewNew}{4}

\newcommand{\nLogTestsFailNewSame}{3}

\newcommand{\nLogRestProjectNewSame}{7}

\newcommand{\nLogWorkflowConfigurationNewSame}{2}

\newcommand{\nLogNetworkOrServiceNewSame}{8}

\newcommand{\nLogRestWorkflowNewSame}{3}

\newcommand{\nLogTestsFailSameNew}{8}

\newcommand{\nLogRestProjectSameNew}{4}

\newcommand{\nLogWorkflowConfigurationSameNew}{3}

\newcommand{\nLogNetworkOrServiceSameNew}{4}

\newcommand{\nLogRestWorkflowSameNew}{1}

\newcommand{\nLogTestsFailSameSame}{3}

\newcommand{\nLogRestProjectSameSame}{3}

\newcommand{\nLogWorkflowConfigurationSameSame}{5}

\newcommand{\nLogNetworkOrServiceSameSame}{7}

\newcommand{\nLogRestWorkflowSameSame}{4}

\newcommand{\nLogTestsFailAll}{21}

\newcommand{\nLogRestProjectAll}{21}

\newcommand{\nLogWorkflowConfigurationAll}{11}

\newcommand{\nLogNetworkOrServiceAll}{22}

\newcommand{\nLogRestWorkflowAll}{12}

\newcommand{\nLogTestsSameCommit}{14}

\newcommand{\nLogTwoPlusRuns}{38}

\newcommand{\nLogSameLastCause}{29}

\newcommand{\pctCompletedRunsSkipped}{23.4}
\newcommand{\pctCompletedRunsCancelled}{3.3}
\newcommand{\pctCompletedRunsActionRequired}{0.1}
\newcommand{\pctWorkflowsNeverFail}{40.8}
\newcommand{\medianRunsWorkflowsAlwaysFail}{5}

\newcommand{\pctRunsFailingFirstFifthThreePlus}{21.6}
\newcommand{\nBreakPairsWithDiff}{469,033}

\newcommand{\pctNeverFailSecurityOnly}{63.5}
\newcommand{\pctNeverFailLicenceOnly}{68.2}
\newcommand{\longestPurposeA}{dependency}
\newcommand{\medianHoursLongestPurposeA}{18.2}
\newcommand{\longestPurposeB}{communication}
\newcommand{\medianHoursLongestPurposeB}{15.8}
\newcommand{\longestPurposeC}{chore}
\newcommand{\medianHoursLongestPurposeC}{14.1}

\newcommand{\pctStartedByScheduleLongestPurposes}{47.1}

\newcommand{\medianHoursOtherPurposes}{6.1}
\newcommand{\pctStartedByScheduleOtherPurposes}{20.5}

\newcommand{\nFailingStreaksFound}{752,873}

\newcommand{\nFailingStreaksDropped}{42,219}

\newcommand{\pctFailingStreaksDropped}{5.6}

\newcommand{\pctFailingTimeLongerThanDayWithDropped}{97.6}

\newcommand{\pctLogProjectWeighted}{44.4}
\newcommand{\pctLogWorkflowWeighted}{43.6}

%% file: Input/introduction.tex
\section{Introduction}\label{sec:introduction}
Continuous Integration and Delivery (CI/CD) is an essential practice in collaborative software development, enabling teams to develop and deliver high-quality software rapidly and reliably. CI/CD workflows automate the process of building, testing, and deploying code changes, allowing developers to detect issues early and ensure that their software is always in a deployable state~\cite{Zampetti2021}.
Many CI/CD platforms provide such automation (\eg GitHub Actions, Travis, Jenkins, GitLab CI, CircleCI). On \github, GitHub Actions (hereafter \gha) has emerged as the dominant CI/CD service~\cite{Golzadeh2022SANER}. %
Repository maintainers can configure \gha workflow files to meet project-specific needs, streamline the development process, and improve characteristics such as efficiency, maintainability, quality, and reliability.

In practice, \gha workflow runs regularly fail~\cite{zheng2025ghafail,GHARuns2026-dataset}. Dealing with such failures increases maintenance effort and reduces the overall reliability of the CI/CD
pipeline~\cite{hilton2016usage,Zampetti2020badpractices}.
Several studies have examined workflow failure dynamics, including the causes of failing runs, reruns, and flaky runs~\cite{zheng2025ghafail,Huang2026GHAReruns, ge2026understandingdetectingflakybuilds,Talebzadeh2026-ICSME}.
These studies analyse failing runs one at a time, or measure the time to recover from a failure, but do not show what starts or ends a failure, nor what happens while the workflow keeps failing.

When a workflow fails, maintainers have to decide whether the last commit broke it, whether the fault lies in the workflow itself or in its environment, whether the failure will go away by itself, and whether anyone is taking care of it. Making such decisions requires looking beyond individual failing runs.
We therefore study uninterrupted series of failing runs of the same workflow, from its first failing run until the workflow succeeds again, referred to as \emph{failing streaks} hereafter.
We analyse the failing streaks of \nRunsInScope{} runs across \nWorkflowsInScope{}
workflows in \nReposInScope{} \github repositories for a period of \windowDays{} days. %

We also consider the fact that not all workflow failures carry the same meaning.
A failing run of a workflow executing a test suite may signal that a genuine regression has been caught, as intended, whereas a failing run of a workflow that merely assigns labels to pull requests may signal a problem with the workflow itself (\eg misconfiguration or insufficient permissions).
To account for this difference, we use an LLM to classify the purposes of each workflow (\eg building, testing, or deploying). This enables us to compare failure and recovery patterns across workflow purposes.
It also lets us distinguish simple workflows that serve a few purposes from more complex workflows that serve many purposes.
To the best of our knowledge, this is the first classification of \gha workflow purposes at this scale.

Our research addresses four research questions:

\begin{description}[leftmargin=!]

    \item[\textbf{\boldmath $RQ_1$}] \textit{How prevalent are failing runs and failing streaks?}
    One run in ten fails, and three in four repositories have at least one failing run. Failing runs rarely come alone: \pctFailingRunsAfterFailingRun{}\% directly follow another failing run. Half of the workflows that can have a failing streak, and two out of three repositories, have at least one failing streak. Workflows with multiple purposes have failing streaks more often than those with a single purpose, even when they exhibit a similar size.

    \item[\textbf{\boldmath $RQ_2$}] \textit{How do failing streaks begin?}
    More than half of the failing streaks start with a \texttt{push} of new commits. The new commits at the start touch the failing workflow file in \pctBreakPairsOwnWorkflowFile{}\% of the cases, against only \pctBaselinePairsOwnWorkflowFile{}\% for commits between two successful runs.
    \pctSameCommitStartAll{}\% %
    of the failing streaks start on the same commit as the last successful run, mostly with triggers other than \texttt{push}, so they are not associated with a new commit.

    \item[\textbf{\boldmath $RQ_3$}] \textit{What happens during a failing streak?}
    Two thirds of the failing streaks have a single failing run, and the median failing streak lasts about two hours. However, the one in five failing streaks that last more than a day account for \pctFailingTimeLongerThanDay{}\% of the time that workflows spend failing. Three in five failing streaks with several failing runs receive new commits while the workflow keeps failing.

    \item[\textbf{\boldmath $RQ_4$}] \textit{How do failing streaks end?}
    Almost all failing streaks end with the trigger that started them, and \pctFailingEndsSameCommit{}\% %
    end on the same commit as the last failing run, without any repository change. When a new commit ends a failing streak, it touches the failing workflow more than three times as often as in normal development. In the logs of 100 sampled failing streaks, those ending on a new commit most often fail because of tests, while those ending on the same commit most often fail because of a network issue or an external service.

\end{description}

%% file: Input/related.tex
\section{Related work}\label{sec:related_work}

Prior studies showed that CI/CD is widely used in open-source projects and changes how these projects are developed~\cite{hilton2016usage,Zhao2017-CI}. Since its public release in 2019, GitHub Actions (\gha) became the dominant CI/CD solution on GitHub, overtaking Travis in less than 18 months~\cite{Golzadeh2022SANER}. Practitioners named the higher \emph{reliability} of \gha as one of the two main reasons for migrating from Travis~\cite{Rostami2023CI}, and rated reliability as the most important characteristic of \gha~\cite{Onsori2026}. This underscores the need for large-scale empirical studies of the extent to which \gha workflows are confronted with reliability issues, such as failing (sequences of) workflow runs.

Many studies have analysed the characteristics of \gha workflow configuration files, such as their structure, complexity, evolution, and usage patterns~\cite{Kinsman2021,decan2022gha,DECAN2023111827,Khatami2024CatchingSmells,ghaleb2025complexity,rostami2026GHAEvolution}, but fewer studies have analysed workflow run failures.
On Travis, failing tests were the main reason for broken builds~\cite{Beller2017,Rausch2017BuildFailures}, and the outcome of a build can itself be misleading, for 9\% of the builds on average~\cite{Gallaba2018NoiseHeterogeneity}.
Zheng~\etal~\cite{zheng2025ghafail} manually analysed 375 failed \gha workflow runs of Java projects: failing tests were again the most common reason of failure (32\%), and most failures (84\%) were caused by the project itself rather than its workflow configuration.
Talebzadeh~\etal~\cite{Talebzadeh2026-ICSME} observed that larger \gha workflows are more likely to exhibit failing runs, a higher downtime, and a longer time to return to a successful run, based on 74K+ runs triggered by workflow file changes.
These studies count failing runs, or measure the time until a workflow passes again, but they do not study what starts a sequence of failing runs, what happens during it, and what ends it.

The failure rate of a \cicd build history only paints a partial picture.
In complement, it is also useful to know whether failures are flaky (\ie they go away on the next rerun without any change to the workflow or code) or persist over consecutive runs. This second category of failures may take more time and effort to resolve.

Rausch~\etal~\cite{Rausch2017BuildFailures} observed that Travis build failures mostly occur in sequence: in ten out of fourteen projects, more than half of the failed builds came right after another failed build.
Gallaba~\etal~\cite{Gallaba2018NoiseHeterogeneity} observed that such sequences have a median length of five failing builds, and last seven days on average before they are fixed.
Jin and Servant~\cite{Jin2020CostEfficientCI} also observed that failures that come right after another failure make up a large part of all failed builds (a median of 52\% per project). They developed SmartBuildSkip, which skips the builds that are predicted to pass, but runs every build after a failure until the build passes again.

Shirakawa~\etal~\cite{shirakawa2026TestAlertSnooze} studied \emph{Test Alert Snooze}, where the same test keeps failing over multiple consecutive commits, in 27 Python and Java projects on \gha. Such snoozes occurred in only four projects, but they covered 42.9\% of the test failures, and the commits made while a test kept failing were rarely bug fixes.
Khatami~\etal~\cite{Khatami2026EASE} studied how developers respond to failing \gha workflow runs, through an in-depth analysis of 21 repositories. They found three response patterns: the most common was \emph{immediate fixing} within minutes or hours; followed by \emph{deferred fixing} a day or more later; and ignoring or abandoning the failure altogether. These patterns were derived from only 21 repositories, so it remains unclear how common they are across \gha. In $RQ_3$, we measure at scale how long failing streaks last, which relates to the first two patterns.

Not every failure needs a code change to resolve it. Huang~\etal~\cite{Huang2026GHAReruns} found that nearly a quarter of 9K+ \gha workflows of Java projects were rerun as a whole at least once, and that rerunning changed the outcome in about a third of these cases, mostly because of flaky tests and external dependencies, at a cost of about 3.7 years of waiting time. Ge~\etal~\cite{ge2026understandingdetectingflakybuilds} found that more than two thirds of the rerun builds of 1,960 Java projects showed flaky behaviour, mostly because of flaky tests, network problems and problems with resolving dependencies.

Although these studies provide valuable insights into the causes and consequences of \cicd failures, none of them studies how sequences of failing runs begin, what happens while they last, and how they end across a large number of \gha workflows and a long time duration.

%% file: Input/methodology.tex
\section{Method}
\label{sec:method}

\subsection{Terminology}\label{sec:terminology}

A \github repository can contain any number of {\bf workflows} to automate tasks such as building, testing, and deploying code.
Workflow configuration files are encoded in YAML format in the {\texttt{.github/workflows}} directory of the repository.
We measure the {\bf size} of a workflow file as its number of {\bf paths}, where a path is an element of the YAML structure of the workflow file~\cite{Talebzadeh2026-ICSME}.
Both measurements are taken from the workflow file at its last run in our observation window.

A {\bf run} corresponds to a single workflow execution, regardless of its outcome.
A run is {\bf triggered} by a specific {\bf trigger event}. %
Typical examples of events are \texttt{push} (a commit is pushed to the branch), \texttt{schedule} (a scheduled time specified with cron syntax in the workflow file), \texttt{workflow\_run} (a run of another workflow starts or completes), \texttt{workflow\_dispatch} (starting the workflow manually), \texttt{issues} (an issue is opened or changed), and \texttt{issue\_comment} (someone comments on an issue or pull request)~\cite{GitHubDocsEvents}.
Each run has a {\bf conclusion} that is either \texttt{success}, \texttt{failure}, \texttt{startup\_failure}, \texttt{skipped}, \texttt{cancelled} or \texttt{action\_required}.
A workflow can specify and run one or more jobs. The {\bf finish time} of a run is the latest \texttt{completed\_at} timestamp among its jobs.

The {\bf run history} of a workflow is the finite sequence of all its observed runs, ordered by start time. Because we focus on failing runs and their transitions to and from successful runs, \emph{we restrict the %
 run history to {\bf successful runs}} \success (\ie~conclusion = \texttt{success}) \emph{and {\bf failing runs}} \failure (\ie~conclusion = \texttt{failure} or \texttt{startup\_failure}) only. Moreover, \emph{we only consider runs on the default branch of the repository} because the focus of the study is on how the mainline of a repository breaks and recovers, %
 and failing streaks only make sense in the context of this mainline.

Given a workflow run history $r$, we define a {\bf run streak} as a contiguous subsequence $s=[ R_i \ldots R_{k} ]$ of $r$, with $1 < i \leq k < length(r)$.
The \textbf{length} of a streak is the number of runs contained in the streak.
A streak of length 1 is called a {\bf single-run streak}, streaks of longer length are called {\bf multi-run streaks}.
A {\bf failing streak} $s$ is a run streak in which the run conclusion transitions from \success to \failure and back:\\
$\text{conclusion}(R_{i-1}) =$~\success,
$\forall R \in s: \text{conclusion}(R) =$~\failure, and
$\text{conclusion}(R_{k+1}) =$~\success.

The \textbf{duration} of a failing streak is computed as $\text{finish}(R_{k+1}) - \text{finish}(R_i)$, \ie the difference between the timestamps at which the workflow history transitions between \success and \failure. When $R_i$ finishes we know that the streak has started because the conclusion has changed to \failure. Similarly, when $R_{k+1}$ finishes we know that the streak has ended because the conclusion has changed to \success.

Each run is executed on a specific {\bf commit} of the repository.
Commits contain file changes (\ie add, remove or modify) that touch either {\bf workflow files} (in the \texttt{.github/workflows} directory), {\bf project files} (\ie all other files of the repository), or both. We refer to {\bf own workflow file} as the workflow file of the workflow run itself.
Commits may change the run conclusion, \ie they may cause a previously successful run to fail, or a previously failing run to succeed.
Consecutive runs in a streak can run on the {\bf same commit}, when no files were pushed to the repository between both runs (\eg for a scheduled or manually triggered run).
The next run in a streak can also be on a {\bf new commit} (\eg when a push event triggers the workflow).
A run can be {\bf re-run} on the same commit (\eg with ``Re-run all jobs'' or ``Re-run failed jobs'' in the \github GUI).

\begin{figure}
  \centering
    \includegraphics[width=\linewidth]{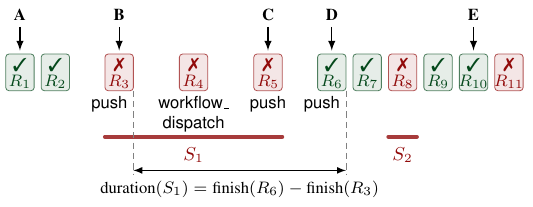}
  \caption{Example of a workflow run history with a failing streak $S_1$ of length 3 and a single-run failing streak $S_2$. An arrow marks each run on a new commit (A to E).
  The run's trigger event is provided below the runs of $S_1$ and its successor run.}
  \label{fig:streaks}
\end{figure}

\fig{fig:streaks} provides a synthetic example of a workflow run history containing 11 runs, including a failing streak $S_1=[R_3,R_4,R_5]$ of length 3 and a single-run failing streak $S_2=[R_8]$.
Failing streak $S_1$ starts with run $R_3$, triggered by a \texttt{push} of a new commit B (the previous runs were on commit A). The streak continues on  commit B with a manual run $R_4$ (trigger \texttt{workflow\_dispatch}) and with another failing run $R_5$ on a new commit C. Successful run $R_6$, triggered by a \texttt{push} of a new commit D, ends the streak. So a new commit may or may not change the outcome of a run.
Note that the analyses hereafter ignore ``incomplete'' failing streaks for which the first or last observed run in the run history is \failure, because we cannot observe their predecessor or successor run to determine the actual length and streak duration.
This excludes \nFailingStreaksDropped{} of the \nFailingStreaksFound{} failing streaks that we found (\pctFailingStreaksDropped{}\%).
For example, %
$[R_{11}]$ in \fig{fig:streaks} is excluded as a failing streak since its successor run falls outside the observed run history.

\subsection{Data Extraction}
\label{sec:dataExtraction}

To study failing streaks, we rely on a recent dataset of \gha workflow-run metadata by Talebzadeh \etal~\cite{GHARuns2026-dataset}, collected from \github-hosted development repositories.
We use version 2026-10-19 of the dataset~\cite{GHARuns2026}, containing \nAllRuns{} runs
of \nAllWorkflows{} workflows in \nAllRepositories{} repositories observed during \windowDaysShort{} days (from \windowStart~to \windowEnd).
The dataset contains all runs of all workflows of its repositories, on all branches and for all trigger events.

We applied the following filters to make it suitable for studying failing streaks of workflow runs:
(1) we only consider runs that are executed on the default branch of the repository (\eg \texttt{main} or \texttt{master} branch), which we retrieved with the \github API for each repository;
(2) we remove runs whose code comes from another repository (a fork), even if its branch has the same name as the default branch;
(3) we remove runs triggered by a pull request (trigger events \texttt{pull\_request} and \texttt{pull\_request\_target}), since they consider proposed code changes rather than the default branch itself;\footnote{For a pull-request run, \github records the branch that the pull request comes from (its head branch), not the branch it merges into (its base branch). A pull request from \texttt{main} into a staging branch therefore produces runs recorded on \texttt{main}, which filter~(1) alone would keep.}
(4) we only keep runs whose conclusion is \success or \failure, where \failure also includes the runs that could not start (\texttt{startup\_failure} represents \pctStartupFailureOfFailingRuns{}\% of the failing runs).

When a run fails, developers can rerun it on GitHub, which creates a new attempt of the same run~\cite{GitHubDocsRerun}. The dataset of~\cite{GHARuns2026-dataset} only contains the last attempt of each run, as returned by the \github API. If the rerun succeeds, the run therefore looks successful, and the failing attempt before it disappears, together with the failing streak that it forms.
We therefore completed the dataset by using the \github REST API to retrieve all earlier attempts of rerun workflows. For each of the \nRunsWithHiddenAttempts{} runs with more than one attempt, we retrieved the conclusion and start time of each attempt, as well as its job timestamps, to compute its finish time. In total, we retrieved \nHiddenAttempts{} earlier attempts. \nHiddenAttemptsInScope{} of those concluded with \success or \failure and were therefore added to the dataset.

\begin{figure}[!t]
  \centering
\begin{tabular}{@{}lc@{}}
    \# runs per workflow & \includegraphics[width=0.6\columnwidth]{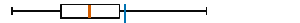} \\
                      & \includegraphics[width=0.6\columnwidth]{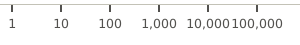} \\
    run duration      & \includegraphics[width=0.6\columnwidth]{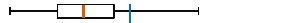} \\
                      & \includegraphics[width=0.6\columnwidth]{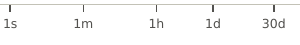} \\
  \end{tabular}
  \caption{Distribution of workflow run characteristics. %
   Box plots use a log scale (orange line =  median; blue line = mean). Run durations below 1 second are drawn at 1 second.}
  \label{fig:overview}
\end{figure}

The above filtering and data completion process resulted in a {\bf dataset of \nRunsInScope{} runs across \nWorkflowsInScope{} workflows related to \nReposInScope{} repositories}.
\fig{fig:overview} reports the distribution (in the form of boxplots in log scale) of the number of runs per workflow and the run duration.
Workflows have a median of \medianRunsPerWorkflow{} runs, while the mean is much higher (\meanRunsPerWorkflow{}). The median run duration is \medianRunSeconds{} seconds, and the mean is again much higher (\meanRunSeconds{} seconds). The large difference between median and mean values signals heavily skewed distributions, in which a few workflows run far more often, and a few runs last far longer, than the others.

To answer $RQ_2$, $RQ_3$ and $RQ_4$ we further extend the dataset with commit information, in order to associate runs to commits that might have caused or fixed a failing run.
For $RQ_2$, we gather the commit associated to the last successful run preceding each failing streak as well as the commit associated to the streak's first failing run.
For $RQ_3$, we gather new commits if consecutive failing runs inside the streak are associated to different commits.
For $RQ_4$, we gather the commit associated to the first successful run of each failing streak as well as the commit associated to the streak's last failing run. This gives \nBreakPairsWithDiff{} pairs of commits at the start of failing streaks, \nDuringPairsWithDiff{} during them, and \nFixPairsWithDiff{} at their end, whose changed files we compare with a baseline.

This baseline is inspired by Rausch~\etal~\cite{Rausch2017BuildFailures} who studied breaking, broken and fixing builds. The baseline is composed of  commits corresponding to successful runs taken from those workflows containing failing streaks.
We took a random sample of 100,000 pairs of consecutive successful runs on different commits.
For each pair, we analysed the files changed between the two commits, which covers all the commits pushed between the two runs (a single push can contain several commits), and classified the changes as affecting only workflow files, %
only project files, %
both, or none (\eg empty commits).
We also recorded whether the changes affected the \emph{own workflow file} (\ie the file of the workflow containing the failing streak).

\subsection{Workflow classification}
\label{sec:classification}
Workflows are used to automate various tasks and serve different \textbf{purposes}, such as building, testing, or deploying code, or handling issues and pull requests~\cite{Onsori2026}.
Depending on its purpose, a failure may be the expected outcome of a workflow.
For example, a workflow that runs a test suite fulfils its job when it fails because it caught a regression, whereas the failure of a workflow that triages issues or publishes a documentation website might signal a problem in the workflow itself.
To compare failing streaks across workflow purposes, we classify each workflow according to the set of purposes it serves.

\paragraph*{Taxonomy and manual labeling}
We started from an initial taxonomy of 15 workflow purposes introduced by Onsori Delicheh \etal~\cite{Onsori2026}. %
We randomly selected 100 workflows from our dataset, and three authors independently assigned purposes from this taxonomy to each workflow by manually reviewing the workflow file contents.
Since a workflow can have several purposes, we measure inter-rater agreement with Krippendorff's $\alpha$~\cite{hayes2007answering}, using the MASI distance between sets of labels.
Krippendorff's $\alpha$ supports multiple labels, any number of raters, and corrects for chance agreement;~the MASI distance gives partial credit to overlapping sets.
The initial multi-author agreement %
was $\alpha=0.53$.
The three raters discussed all disagreements and agreed on a final set of purposes for each workflow.
During this consolidation phase, the \textit{PR management} and \textit{issue management} purposes were merged into a single \textit{pr-and-issues} purpose, and two missing purposes were added to  the taxonomy:
\textit{chore}, for repository housekeeping (\eg deleting stale branches, cleaning old artefacts);
and \textit{workflow}, for workflows that delegate their work to other workflows, either by triggering them (\eg with \texttt{gh workflow run}) or by calling a reusable workflow.
\tab{tab:purposes} lists the consolidated taxonomy used in this paper.

\begin{table}[!t]
  \caption{Workflow purposes in our dataset: \% workflows having this purpose and \% workflows having it as their only purpose.}
  \label{tab:purposes}
  \centering
  \footnotesize
  \setlength{\tabcolsep}{3pt}
  \begin{tabular}{@{}llrr@{}}
    Purpose        & Description                            &   \% & only (\%) \\
    \midrule
    build          & Compile, build, or package artefacts   & 45.6 &  5.1 \\
    test           & Run tests, benchmarks, etc.            & 43.3 & 11.8 \\
    quality        & Lint, analyse, or review code          & 27.9 & 19.1 \\
    release        & Version and publish distributables     & 23.4 &  7.7 \\
    report         & Publish results (\eg coverage), badges & 17.9 &  2.0 \\
    pr-and-issues  & Act on pull requests and issues        & 16.1 & 35.4 \\
    docs           & Build or publish documentation         & 12.7 & 36.1 \\
    workflow       & Trigger or reuse other workflows       & 11.3 &  6.9 \\
    security       & Scan for vulnerabilities, secrets, etc.&  8.9 & 41.9 \\
    dependency     & Update, pin, find unused dependencies  &  4.5 &  4.3 \\
    chore          & Repository housekeeping                &  4.4 & 30.3 \\
    communication  & Notify outside \github (\eg Slack)     &  2.8 & 13.4 \\
    deploy         & Deploy to a runtime environment        &  1.7 & 13.0 \\
    licence        & Check licence or CLA compliance        &  1.5 & 17.0 \\
    infrastructure & Provision infrastructure               &  0.6 & 19.1 \\
    other          & None of the above                      &  0.8 & 100.0 \\
  \end{tabular}
\end{table}

\paragraph*{Automated classification}
To scale the classification to all workflows, we used OpenAI's \texttt{gpt-6-luna} model, with a medium reasoning effort, to classify each workflow according to the consolidated taxonomy.
The prompt contains the taxonomy with a description of each purpose, the contents of the workflow file to label, and labeling instructions.
The LLM is tasked to assign every purpose for which a concrete job, step, or command performs the corresponding task, ignoring steps that only support other tasks (\eg checkout, toolchain setup, or caching).
We ran Luna on the 100 manually classified workflows and compared the inferred purposes with our consolidated ones.
Considering Luna as an additional rater, the agreement between Luna and our consolidated labels is $\alpha=0.76$.
Luna assigns 179 of our 187 purposes (recall of 0.96), and 179 of the 219 purposes it assigns are ours (precision of 0.82).
Luna tends to assign more purposes than we did, in particular \emph{workflow} (14 times, against 3).
A second look at the 32 workflows with disagreements revealed that, for 18 of them, the purpose set of Luna was the correct one and our consolidated labels had missed a purpose (\eg a call to a reusable workflow).
Another independent evaluation following the same protocol with \texttt{gpt-6.1-sol} as the rater yielded similar results.
This made us confident that Luna can label all workflows in our dataset.

\paragraph*{Classification results}
We classified each workflow file at its last run in the observation window (\sect{sec:terminology}).
Luna returned a valid set of purposes for all of them.
Workflows have a mean of 2.23 purposes (median 2, maximum 13): 33.0\% have a single purpose, 34.0\% have two, 18.5\% have three, and 14.5\% have four or more.
\emph{build} (45.6\%) and \emph{test} (43.3\%) are by far the most common purposes, and \{\emph{build}, \emph{release}\} and \{\emph{build}, \emph{test}\} are the most common purpose sets (7.3\% and 7.2\%).
Some purposes are rarely the only purpose of a workflow, such as \emph{build} (5.1\%) and \emph{report} (2.0\%), whereas others are often served by a dedicated workflow, such as \emph{security} (41.9\%), \emph{docs} (36.1\%), and \emph{pr-and-issues} (35.4\%).
Only 0.8\% of the workflows fit none of the purposes of the taxonomy.

%% file: Input/rq1.tex
\section{\texorpdfstring{$RQ_1$}{RQ1}: How prevalent are failing runs and failing streaks?}\label{sec:rq1}

To motivate the study of failing streaks, we first assess how often runs fail and whether failures occur alone or in streaks.

\paragraph*{Failing runs}

\pctRunsFailing{}\% of all runs in the dataset %
fail. They belong to \pctWorkflowsWithFailingRun{}\% of the workflows and \pctReposWithFailures{}\% of the repositories. These workflows with failing runs have a median run history length of \medianRunsWorkflowsWithFailingRun{}.
\pctWorkflowsNeverFail{}\% of the workflows {\bf never} fail, have shorter run histories than the workflows with failing runs (median length of \medianRunsWorkflowsNeverFail{}), and together account for only \pctRunsInWorkflowsNeverFail{}\% of all runs. 
Most workflows whose only purpose is \emph{security} (\pctNeverFailSecurityOnly{}\%) or \emph{licence} (\pctNeverFailLicenceOnly{}\%) never fail, possibly because they inspect the repository and provide reports rather than build or run its code.

\pctWorkflowsAlwaysFail{}\% of all workflows fail on {\bf every} run, but have short run histories (median length of \medianRunsWorkflowsAlwaysFail{}). Most failing runs do not occur alone: \pctFailingRunsAfterFailingRun{}\% directly follow another failing run of the same workflow.
This shows why studying individual failing runs is not sufficient. A failing run may simply repeat an earlier failure, and the number of failing runs depends on how often a workflow runs while it remains broken.
This motivated us to study and quantify the failing streaks.

\paragraph*{Failing streaks}
A failing streak requires a run history containing at least one failing run and two successful runs (one before and one after the failure), so the \nWorkflowsOneOrTwoRuns{} workflows with fewer than three runs cannot have failing streaks. %
Of the remaining \nWorkflowsCanHaveStreak{} workflows, \nFailingStreaksPerWorkflow{} (\pctWorkflowsWithFailingStreak{}\%) have at least one failing streak, as do \pctReposWithFailingStreak{}\% of the repositories containing them.
Our dataset contains \nFailingStreaks{} failing streaks in total.
A workflow with failing streaks has a median of \medianFailingStreaksPerWorkflow{} of them. Some workflows with a failing run have no failing streak, because they never succeeded during our observation window, were already failing when the window began, or were still failing when it ended.

\begin{table}[!t]
  \caption{Failing streaks of the workflows, grouped by run history length.}
  \label{tab:streaks-by-runs}
  \centering
  \begin{tabular}{@{}lrrr@{}}
    \# runs & \# workflows & {\shortstack{with a\\ failing streak (\%)}} & {\shortstack{\# failing streaks\\(median)}} \\
    \midrule
    \minRunsFirstFifth{}--2    $^*$                   & \nWorkflowsOneOrTwoRuns{} & 0.0                          & 0 \\
    3--\maxRunsFirstFifth{}                      & \nWorkflowsFirstFifth{}   & \pctWithFailingStreakFirstFifth{}  & \medianFailingStreaksFirstFifth{}  \\
    \minRunsSecondFifth{}--\maxRunsSecondFifth{} & \nWorkflowsSecondFifth{}  & \pctWithFailingStreakSecondFifth{} & \medianFailingStreaksSecondFifth{} \\
    \minRunsThirdFifth{}--\maxRunsThirdFifth{}   & \nWorkflowsThirdFifth{}   & \pctWithFailingStreakThirdFifth{}  & \medianFailingStreaksThirdFifth{}  \\
    \minRunsFourthFifth{}--\maxRunsFourthFifth{} & \nWorkflowsFourthFifth{}  & \pctWithFailingStreakFourthFifth{} & \medianFailingStreaksFourthFifth{} \\
    \minRunsLastFifth{}+                         & \nWorkflowsLastFifth{}    & \pctWithFailingStreakLastFifth{}   & \medianFailingStreaksLastFifth{}   \\
    \midrule
    all with 3+ runs                             & \nWorkflowsAllWorkflows{} & \pctWithFailingStreakAllWorkflows{} & \medianFailingStreaksAllWorkflows{} \\
    \multicolumn{4}{@{}l}{\footnotesize $^*$A failing streak needs at least three runs.} \\
  \end{tabular}
\end{table}

To understand the relationship between the number of observed runs and the number of failing streaks, we group the workflows in \tab{tab:streaks-by-runs} by run history length.
Since more runs imply more chances to fail, both the proportion of workflows with at least one failing streak and the median number of failing streaks increase with the number of observed runs: from \pctWithFailingStreakFirstFifth{}\% (median \medianFailingStreaksFirstFifth{}) for the group of 3--\maxRunsFirstFifth{} runs to \pctWithFailingStreakLastFifth{}\% (median \medianFailingStreaksLastFifth{}) for the group of \minRunsLastFifth{}+ runs.
The median number of failing streaks grows much more slowly than the number of runs, for two reasons.
First, workflows that run more often fail on a smaller proportion of their runs (from \pctRunsFailingFirstFifthThreePlus{}\% of runs for workflows with 3--\maxRunsFirstFifth{} runs to \pctRunsFailingLastFifth{}\% for those with \minRunsLastFifth{}+ runs), which aligns with Khatami~\etal~\cite{Khatami2026EASE}. Second, when they fail, their failing streaks contain more failing runs (a mean of \meanFailingRunsPerStreakFirstFifth{} for workflows with 3--\maxRunsFirstFifth{} runs, compared to \meanFailingRunsPerStreakLastFifth{} for those with \minRunsLastFifth{}+ runs).

\summary{One run in ten fails, and three in four failing runs directly follow another failing run of the same workflow. Half of the workflows with at least three runs, and two in three of their repositories, have at least one failing streak.}

\begin{table}[!t]
\caption{Failing streaks by number of purposes, for the \nWorkflowsSizeTable{} workflows with at least three runs and a purpose label, with the three most common purpose sets of each group. The median counts only the workflows with a failing streak.}

  \label{tab:purpose-sets}
  \centering
  \footnotesize
  \setlength{\tabcolsep}{3.5pt}
  \begin{tabular}{@{}lrrr@{}}
    purpose set & {\shortstack{\# workflows}} & {\shortstack{with a failing\\streak (\%)}} & {\shortstack{failing streaks\\(median)}} \\
    \midrule
    \textbf{1 purpose}           & \nWorkflowsPurposesOne{}          & \pctWithStreakPurposesOne{}          & \medianStreaksPurposesOne{} \\
    \quad \purposeSetOneA{}      & \nWorkflowsPurposeSetOneA{}       & \pctWithStreakPurposeSetOneA{}       & \medianStreaksPurposeSetOneA{} \\
    \quad \purposeSetOneB{}      & \nWorkflowsPurposeSetOneB{}       & \pctWithStreakPurposeSetOneB{}       & \medianStreaksPurposeSetOneB{} \\
    \quad \purposeSetOneC{}      & \nWorkflowsPurposeSetOneC{}       & \pctWithStreakPurposeSetOneC{}       & \medianStreaksPurposeSetOneC{} \\
    \midrule
    \textbf{2 purposes}          & \nWorkflowsPurposesTwo{}          & \pctWithStreakPurposesTwo{}          & \medianStreaksPurposesTwo{} \\
    \quad \purposeSetTwoA{}      & \nWorkflowsPurposeSetTwoA{}       & \pctWithStreakPurposeSetTwoA{}       & \medianStreaksPurposeSetTwoA{} \\
    \quad \purposeSetTwoB{}      & \nWorkflowsPurposeSetTwoB{}       & \pctWithStreakPurposeSetTwoB{}       & \medianStreaksPurposeSetTwoB{} \\
    \quad \purposeSetTwoC{}      & \nWorkflowsPurposeSetTwoC{}       & \pctWithStreakPurposeSetTwoC{}       & \medianStreaksPurposeSetTwoC{} \\
    \midrule
    \textbf{3 purposes}          & \nWorkflowsPurposesThree{}        & \pctWithStreakPurposesThree{}        & \medianStreaksPurposesThree{} \\
    \quad \purposeSetThreeA{}    & \nWorkflowsPurposeSetThreeA{}     & \pctWithStreakPurposeSetThreeA{}     & \medianStreaksPurposeSetThreeA{} \\
    \quad \purposeSetThreeB{}    & \nWorkflowsPurposeSetThreeB{}     & \pctWithStreakPurposeSetThreeB{}     & \medianStreaksPurposeSetThreeB{} \\
    \quad \purposeSetThreeC{}    & \nWorkflowsPurposeSetThreeC{}     & \pctWithStreakPurposeSetThreeC{}     & \medianStreaksPurposeSetThreeC{} \\
    \midrule
    \textbf{4+ purposes}         & \nWorkflowsPurposesFourPlus{}     & \pctWithStreakPurposesFourPlus{}     & \medianStreaksPurposesFourPlus{} \\
    \quad \purposeSetFourPlusA{} & \nWorkflowsPurposeSetFourPlusA{}  & \pctWithStreakPurposeSetFourPlusA{}  & \medianStreaksPurposeSetFourPlusA{} \\
    \quad \purposeSetFourPlusB{} & \nWorkflowsPurposeSetFourPlusB{}  & \pctWithStreakPurposeSetFourPlusB{}  & \medianStreaksPurposeSetFourPlusB{} \\
    \quad \purposeSetFourPlusC{} & \nWorkflowsPurposeSetFourPlusC{}  & \pctWithStreakPurposeSetFourPlusC{}  & \medianStreaksPurposeSetFourPlusC{} \\
  \end{tabular}
\end{table}

\paragraph*{Workflow purposes}
Finally, we compare how prevalent failing streaks are across workflows with different purposes.
\tab{tab:purpose-sets} reveals that workflows serving multiple purposes have failing streaks more often.
More than a third of the single-purpose workflows (\pctWithStreakPurposesOne{}\%) have at least one failing streak, with a median of \medianStreaksPurposesOne{}.
More than two-thirds of the workflows that serve four or more purposes (\pctWithStreakPurposesFourPlus{}\%) have at least one failing streak, with a median of \medianStreaksPurposesFourPlus{}.
This difference is not explained by how often the workflows run: within each run history length group of \tab{tab:streaks-by-runs}, the proportion of workflows with a failing streak still increases with the number of purposes.
Workflow size (measured as the number of paths in the workflow file; see \sect{sec:terminology}) does not explain the difference either. Talebzadeh~\etal~\cite{Talebzadeh2026-ICSME} found that larger workflows tend to fail more often. Workflows with more purposes are generally larger (a median of \medianPathsOneLabel{} paths for one purpose, compared with \medianPathsFourPlusLabels{} for four or more; Spearman's $\rho=\spearmanLabelsPaths{}$). However, when splitting workflows into three equally sized groups, the share with a failing streak still increases with the number of purposes in each group (\tab{tab:purpose-size}).
The same observation holds when restricting the analysis to workflows that run mainly on \texttt{push} (\pctSizeTableMainTriggerPush{}\% of them) or mainly on \texttt{schedule} (\pctSizeTableMainTriggerSchedule{}\%): within the same size group, workflows with more purposes break more often.

As \tab{tab:purpose-sets} also shows, single-purpose \emph{test} workflows have a higher proportion of failing streaks than those involving other purposes. For example, \pctWithStreakPurposeSetOneC{}\% of \emph{test} workflows have a failing streak, compared with \pctWithStreakPurposeSetOneA{}\% of \emph{pr-and-issues} workflows. This is plausible because test workflows are expected to fail when a change breaks a test. As the log analysis in $RQ_4$ will show, such failures often point to a problem in the code rather than in the workflow.

\begin{table}[!t]
  \caption{\% workflows with a failing streak, by size group (number of paths) and number of purposes.}
  \label{tab:purpose-size}
  \centering
  \footnotesize
  \setlength{\tabcolsep}{3.5pt}
  \begin{tabular}{@{}lrrrr@{}}
     & \multicolumn{4}{c}{number of purposes of the workflow} \\
    \cmidrule(l){2-5}
    workflow size (\#paths) & 1 & 2 & 3 & 4+ \\
    \midrule
    small ($\leq$\maxPathsSmall{})                & \pctWithStreakSmallOneLabel{}  & \pctWithStreakSmallTwoLabels{}  & \pctWithStreakSmallThreeLabels{}  & \pctWithStreakSmallFourPlusLabels{} \\
    medium (\minPathsMedium{}--\maxPathsMedium{}) & \pctWithStreakMediumOneLabel{} & \pctWithStreakMediumTwoLabels{} & \pctWithStreakMediumThreeLabels{} & \pctWithStreakMediumFourPlusLabels{} \\
    large (\minPathsLarge{}+)                     & \pctWithStreakLargeOneLabel{}  & \pctWithStreakLargeTwoLabels{}  & \pctWithStreakLargeThreeLabels{}  & \pctWithStreakLargeFourPlusLabels{} \\
  \end{tabular}
\end{table}

\summary{Workflows with more purposes have failing streaks more often, even when they share a similar size.}

%% file: Input/rq2.tex
\section{\texorpdfstring{$RQ_2$}{RQ2}: How do failing streaks begin?}\label{sec:rq2}

$RQ_1$ showed that failing streaks are common.
$RQ_2$ aims to explore how they begin by focusing on the trigger event, commit and workflow purposes associated with each streak's first failing run.

\paragraph*{Trigger analysis} First, we identify the commit associated with the first failing run of each failing streak and its trigger, to determine whether the streak started on a new commit or on the same commit as the last successful run.
Most failing streaks (\pctDifferentCommitStartAll{}\%) start on a \emph{new} commit, different from that of the previous successful run, so one or more commits came between the two runs, which may have caused the failure.
The other \pctSameCommitStartAll{}\% start without any change to the repository, so their causes may lie outside it, for example in flaky builds or tests, unavailable external services, or incompatible external dependencies.

\begin{table}[!t]
\caption{Characteristics of failing streaks, grouped by trigger of the first run of the failing streak.}
  \label{tab:streaks-starting-triggers}
  \centering
  \begin{tabular}{@{}lrrrr@{}}
    {trigger} & {\shortstack{\# failing\\ streaks}} & {\shortstack{\% failing\\streaks}} & {\shortstack{\% streaks starting\\ on same commit}} \\
    \midrule
    \texttt{push}                & \nFailingStreaksStartedByPush{}             & \pctFailingStreaksStartedByPush{}             & \pctSameCommitStartPush{}             \\
    \texttt{schedule}            & \nFailingStreaksStartedBySchedule{}         & \pctFailingStreaksStartedBySchedule{}         & \pctSameCommitStartSchedule{}         \\
    \texttt{workflow\_run}       & \nFailingStreaksStartedByWorkflowRun{}      & \pctFailingStreaksStartedByWorkflowRun{}      & \pctSameCommitStartWorkflowRun{}      \\
    \texttt{workflow\_dispatch}  & \nFailingStreaksStartedByWorkflowDispatch{} & \pctFailingStreaksStartedByWorkflowDispatch{} & \pctSameCommitStartWorkflowDispatch{} \\
    other               & \nFailingStreaksStartedByOther{}            & \pctFailingStreaksStartedByOther{}            & \pctSameCommitStartOther{}            \\
    \midrule
    all                 & \nFailingStreaks{}                          & 100.0                                         & \pctSameCommitStartAll{}              \\
  \end{tabular}
\end{table}

\tab{tab:streaks-starting-triggers} reports the triggers that start failing streaks, together with  the number and proportion of failing streaks whose first run it triggered, and the proportion of these streaks that start on the same commit as the previous successful run.
The most common trigger by far is \texttt{push}, accounting for \pctFailingStreaksStartedByPush{}\% of the first runs of failing streaks.
\texttt{push} nearly always starts a failing streak on a different commit than the previous run, which is expected since a push event corresponds to new commits.

Other popular triggers are \texttt{schedule} (\pctFailingStreaksStartedBySchedule{}\%), \texttt{workflow\_run} (\pctFailingStreaksStartedByWorkflowRun{}\%), and \texttt{workflow\_dispatch} (\pctFailingStreaksStartedByWorkflowDispatch{}\%). All remaining triggers bundled together start only \pctFailingStreaksStartedByOther{}\% of the failing streaks.
All triggers other than \texttt{push} start most of their failing streaks on the same commit as the last successful run, from \pctSameCommitStartWorkflowDispatch{}\% for \texttt{workflow\_dispatch} and \pctSameCommitStartSchedule{}\% for \texttt{schedule} to \pctSameCommitStartWorkflowRun{}\% for \texttt{workflow\_run}. These triggers are therefore more likely to start failing streaks without new commits pushed to the repository, \ie without any change to the repository, which points to causes outside the repository or to flaky failures.

\summary{More than half of the failing streaks start with a \texttt{push} of a new commit, but one in three starts on the same commit as the last successful run, almost always from a trigger other than a push, so their failure does not come from a new commit.}

\paragraph*{Commit analysis}
Next, for the failing streaks that start on a new commit, we study which files changed between the last successful run and the first failing run: workflow files, other project files, or both.
To do so, we compare these pairs of $\langle$commit, failing run$\rangle$ against the baseline of pairs of consecutive successful runs on different commits (Section~\ref{sec:dataExtraction}).
Most failing streaks start with a change to project files only (\pctBreakPairsProjectFilesOnly{}\%, compared to \pctBaselinePairsProjectFilesOnly{}\% in the baseline).
Commits that change both project and workflow files are more common at the start of a failing streak (\pctBreakPairsWorkflowAndProjectFiles{}\%, compared to \pctBaselinePairsWorkflowAndProjectFiles{}\% in the baseline).
The changes at the start of \pctBreakPairsOwnWorkflowFile{}\% of these failing streaks touch the file defining the failing workflow itself, more than twice as often as in the baseline (\pctBaselinePairsOwnWorkflowFile{}\%). Still, most failing streaks start without any change to the workflow itself, so a failing streak usually starts from a change in the project or in its environment, rather than in the workflow configuration.

\summary{Most commits that start a failing streak change project files only, but they change the file of the failing workflow itself more than twice as often as in normal development.}

\paragraph*{Workflow purposes}
Finally, we examine whether the number of purposes of a workflow (\sect{sec:classification}) is associated with how its failing streaks begin. Workflows with more purposes start a larger share of their failing streaks with a \texttt{push} than workflows with fewer purposes. However, among failing streaks started by the same trigger (for example, \texttt{push}), the number of purposes makes little difference: about the same share starts without a new commit, and new commits change similar kinds of files. %

\summary{How a failing streak begins goes with its trigger rather than with the number of purposes of the workflow.}

%% file: Input/rq3.tex
\section{\texorpdfstring{$RQ_3$}{RQ3}: What happens during a failing streak?}\label{sec:rq3}

While $RQ_2$ analysed how failing streaks begin, $RQ_3$ studies what happens while a workflow keeps failing. Ideally, a workflow that fails should succeed upon its next run, so that failing streaks are single-run streaks at best. Multi-run failing streaks cost more compute time and maintainers' effort, and while a workflow keeps failing, it can hide new problems, so a new bug pushed during the failing streak stays hidden behind the ongoing failure~\cite{shirakawa2026TestAlertSnooze}.

\begin{figure}[!t]
  \centering
  \includegraphics[width=.8\columnwidth]{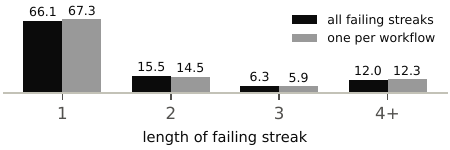}
  \caption{Share of failing streaks (\%) containing 1, 2, 3, or 4+ failing runs, for all failing streaks and for one random failing streak per workflow.}
  \label{fig:failing-runs}
\end{figure}

\begin{figure}[!t]
  \centering
  \small
  \setlength{\tabcolsep}{4pt}
  \begin{tabular}{@{}lc@{}}
    all failing streaks (\nFailingStreaks{})          & \includegraphics[width=0.55\columnwidth]{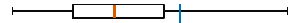} \\
    one per workflow (\nFailingStreaksPerWorkflow{})  & \includegraphics[width=0.55\columnwidth]{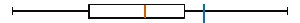} \\
                                                      & \includegraphics[width=0.55\columnwidth]{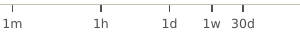} \\
  \end{tabular}
  \caption{Distribution of the duration for all failing streaks, and for one random failing streak per workflow. Box plots in log scale, median = orange line, mean = blue line.}
  \label{fig:duration-streaks}
\end{figure}

\paragraph*{Length and duration}
We start by analysing the length (\fig{fig:failing-runs}) and duration (\fig{fig:duration-streaks}) of failing streaks. We do this for all failing streaks, as well as for a subset of one random failing streak per workflow, in order to avoid possible bias induced by workflows that contain a high number of failing streaks.
\fig{fig:failing-runs} shows that most failing streaks have few failing runs: about two thirds %
are single-run streaks (\ie the next run already succeeds), and only \pctLengthFourPlus{}\% (resp.\ \pctLengthFourPlusPerWorkflow{}\%) have four or more runs.

The median duration of a failing streak is \medianFailingHours{} hours (or \medianFailingHoursPerWorkflow{} hours when we randomly select only one failing streak per workflow).
This low median is mainly due to the single-run failing streaks that have a median duration of \medianHoursOneRun{} hour, whereas failing streaks of length 2 or more have a median duration of \medianHoursTwoPlusRuns{} hours.
Note that the duration also depends on how often a workflow runs, for example if the workflow is scheduled to run once a day, it cannot recover in less than a day, even if the fix takes only minutes.

The time that workflows spend failing is very skewed. If we add up the durations of all failing streaks of all workflows, only \pctStreaksLongerThanDay{}\% of all failing streaks have a duration of more than a day, but they make up \pctFailingTimeLongerThanDay{}\% of this total time. Only \pctStreaksLongerThanWeek{}\% of all failing streaks last more than a week, but still account for \pctFailingTimeLongerThanWeek{}\% of this total time.

\summary{Two thirds of the failing streaks end at the next run, but the one in five that last more than a day make up more than 90\% of the time that workflows spend failing.}

\paragraph*{Trigger analysis} Next, we study the run triggers of failing streaks,  focusing on the subset of \nStreaksWithPairAll{} multi-run failing streaks.
If a run in a failing streak has a different trigger than the failing run that initiated the streak, this could signal attempts to fix the failure.
However, \pctOneEventFailingStreak{}\% of the multi-run failing streaks keep the same trigger for all their  runs. For example, a workflow that fails on a schedule keeps failing on its schedule, and a workflow that fails on a push keeps failing on the next pushes. So a change of trigger, such as a manual run, rarely shows a reaction to the failure.

\paragraph*{Commit analysis} We also analyse whether new commits are made during a multi-run failing streak. These commits could be attempts to fix the failure, either in the workflow file or in the project (\eg a failing test). %
As in $RQ_2$, we compare pairs of consecutive failing runs on different commits against the baseline of consecutive successful runs on different commits.
\pctStreaksNewCommitWhileFailing{}\% of the multi-run failing streaks receive at least one new commit between two consecutive runs.
Most of these commits change project files only (\pctDuringPairsProjectFilesOnly{}\%, against \pctBaselinePairsProjectFilesOnly{}\% in the baseline), \pctDuringPairsWorkflowAndProjectFiles{}\% change both project and workflow files (against \pctBaselinePairsWorkflowAndProjectFiles{}\%), and the failing workflow itself is touched in \pctDuringPairsOwnWorkflowFile{}\% of these commits, about twice as often as in the baseline (\pctBaselinePairsOwnWorkflowFile{}\%). So development goes on while a workflow is failing, and the more frequent changes to the workflow's own file suggest that some of these commits try to fix it.
Development goes on even in the scheduled failing streaks that last more than a week: \pctNewCommitLongSchedule{}\% of them receive new commits while the workflow keeps failing. So the project is not abandoned, yet the scheduled workflow stays broken, whether or not these commits try to fix it.
A possible reason is that the commit authors are not even aware that the workflow is broken: \github sends notifications of a scheduled workflow only to the author who last changed its cron schedule~\cite{GitHubDocsNotifications}, and only if this author has opted in to them~\cite{GitHubDocsManageNotifications}.

\summary{Three in five failing streaks with several failing runs receive new commits while the workflow keeps failing, and so do three quarters of the scheduled failing streaks that last more than a week.}

\paragraph*{Workflow purposes}
We checked whether workflow purposes affect how long failing streaks last by comparing single-purpose workflows with workflows that have four or more purposes. Across all failing streaks, single-purpose workflows recover faster (a median of \medianHoursOnePurpose{} hours, against \medianHoursFourPlusPurposes{} hours). However, this comes from a small group of workflows: the 1\% of single-purpose workflows with the most failing streaks (\nWorkflowsTopOnePctOnePurpose{} workflows) produce \pctStreaksTopOnePctOnePurpose{}\% of their failing streaks, which last a median of \medianMinutesTopOnePctOnePurpose{} minutes. With one random failing streak per workflow, both groups stay in the failure state about equally long (a median of \medianHoursOnePurposePerWorkflow{} and \medianHoursFourPlusPurposesPerWorkflow{} hours).
Workflows whose only purpose is \emph{\longestPurposeA}, \emph{\longestPurposeB}, or \emph{\longestPurposeC} stay failing longest: with one random failing streak per workflow, they stay failing for a median of \medianHoursLongestPurposeA{}, \medianHoursLongestPurposeB{}, and \medianHoursLongestPurposeC{} hours, against \medianHoursOtherPurposes{} hours for workflows with any other single purpose. This is consistent with their trigger where \pctStartedByScheduleLongestPurposes{}\% of these failing streaks start with a \texttt{schedule} (against \pctStartedByScheduleOtherPurposes{}\%). A scheduled workflow only runs again at its next scheduled time, so it cannot pass sooner even if the cause is fixed quickly. At the same trigger, they last as long as the other workflows.

%% file: Input/rq4.tex
\section{\texorpdfstring{$RQ_4$}{RQ4}: How do failing streaks end?}\label{sec:rq4}

Having looked at how failing streaks start and what happens during them, we now analyse how they end, from three different viewpoints:
(1) which trigger ends a failing streak and whether it ends on a new commit, compared to how the streak started;
(2) for the failing streaks that end on a new commit, whether this commit touched the workflow file or other project files; and
(3) whether a failing streak starts and ends on a new commit or on the same commit, and, for a sample of failing streaks, what caused the failure.

First, we find that failing streaks almost always end with the trigger that started them (\pctFailingStreaksSameTriggerBothEnds{}\%); for example, only \pctScheduleStartEndedByPush{}\% of the failing streaks started by a \texttt{schedule} end with a \texttt{push}. A \texttt{push} ends \pctFailingStreaksEndedByPush{}\% of the failing streaks. Of these, only \pctSameCommitEndPush{}\% end on the same commit, almost always after a rerun of the failed run. On the other hand, the other triggers end most of their failing streaks on the same commit as the last failing run, from \pctSameCommitEndSchedule{}\% for \texttt{schedule} to \pctSameCommitEndWorkflowRun{}\% for \texttt{workflow\_run}. Overall, \pctSameCommitEndAll{}\% of the failing streaks end on the same commit, while \pctSameCommitStartAll{}\% start on the same commit.

\summary{Failing streaks almost always end with the trigger that started them, and two in five end on the same commit as the last failing run, without any change to the repository.}

Second, we consider the failing streaks that end on a new commit, and compare the files changed between the last failing run and the first successful run against the same baseline as in $RQ_2$. \pctFixPairsProjectFilesOnly{}\% of these commits change project files only (against \pctBaselinePairsProjectFilesOnly{}\% in the baseline), and \pctFixPairsWorkflowAndProjectFiles{}\% change both project and workflow files (against \pctBaselinePairsWorkflowAndProjectFiles{}\%). The failing workflow itself is changed in \pctFixPairsOwnWorkflowFile{}\% of these commits, more than three times as often as in the baseline (\pctBaselinePairsOwnWorkflowFile{}\%), and more often than at the start (\pctBreakPairsOwnWorkflowFile{}\%) or during a failing streak (\pctDuringPairsOwnWorkflowFile{}\%).

\summary{%
 When a failing streak ends on a new commit, this commit most often changes project files only, but it changes the failing workflow itself more than three times as often as in normal development.}

\begin{figure}[!h]
    \centering
    \includegraphics[width=.9\linewidth]{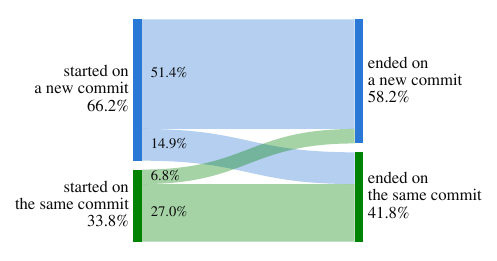}
    \caption{The four groups of failing streaks. The commit at their start (left) and at their end (right) is either a new commit or the same commit as the run before. Each band shows its share of all failing streaks.}
    \label{fig:streaks-start-end-groups}
\end{figure}

Third, we look further into how failing streaks end, by relating how they start to how they end. A failing streak can start on a new commit or on the same commit as the last successful run, and it can end on a new commit or on the same commit as the last failing run. This gives four groups of failing streaks, shown in %
\fig{fig:streaks-start-end-groups}.
About half of the failing streaks (\pctStartedOnDiffCommitEndedOnDiffCommit{}\%) start and end on a new commit.
More than a quarter (\pctStartedOnSameCommitEndedOnSameCommit{}\%) start and end on the same commit, so they do not coincide with changes to the repository. These streaks are the shortest (a median of \medianHoursStartedOnSameCommitEndedOnSameCommit{} hours).
\pctStartedOnDiffCommitEndedOnSameCommit{}\% of the failing streaks start on a new commit but end on the same commit.
Finally, \pctStartedOnSameCommitEndedOnDiffCommit{}\% start on the same commit but end on a new commit. They last the longest (a median of \medianHoursStartedOnSameCommitEndedOnDiffCommit{} hours, with \pctLongerThanDayStartedOnSameCommitEndedOnDiffCommit{}\% above 24 hours), which suggests that an external change broke the workflow and that a new commit was needed to fix it.
We observe similar proportions between single-run and multi-run failing streaks.
Overall, about two thirds of the failing streaks start on a new commit. However, \pctSameCommitEndAll{}\% of all failing streaks eventually recover on the same commit, so the failure might be due to external factors (\eg missing permissions).
Conversely, a failing streak ending on a new commit does not necessarily imply that this commit is a direct fix to the failure.

To understand what causes failures and how workflows recover, we inspect the logs of a sample of failing streaks.
We randomly draw 25 failing streaks from each of the four groups (100 in total), selecting at most one failing streak from each repository in each group, among those with available logs (the retention period is 90 days~\cite{GitHubDocsRetention}).
Two authors analysed 50 of these streaks each, following the same methodology. For each failing streak, they identified (1) the cause of the first failing run, using the failure categories of Zheng~\etal~\cite{zheng2025ghafail}, and (2) the cause of the last failing run for multi-run failing streaks. %
To compare logs, they used CiDiff~\cite{Hubner2025CiDiff} to inspect the diff between two logs.
For (1), they inspected the diff between the logs of the first failing run of the streak and its previous successful run. For (2), they inspected the diff between the logs of the last failing run of the streak and its next successful run.

\tab{tab:log-causes} shows the cause of the first failing run of each sampled failing streak. The last run failed for the same cause as the first one in \nLogSameLastCause{} of the \nLogTwoPlusRuns{} multi-run failing streaks, so a failing streak usually fails for the same reason from its first to its last failing run.
When a failing streak ends on a new commit, the most common cause is failing tests (\nLogTestsFailNewNew{} and \nLogTestsFailSameNew{} of 25 in these two groups). When it ends on the same commit, the workflow or its infrastructure causes most failing streaks (\nLogWorkflowNewSame{} and \nLogWorkflowSameSame{} of 25), most often due to a network issue or an external service failure. Failing tests also appear in the three groups where the same commit both passed and failed (\nLogTestsSameCommit{} failing streaks): these tests failed without any change to the repository, which points to flaky tests. %
Workflows that build or test code make up most of the first three groups (\nLogBuildTestSameNew{} to \nLogBuildTestNewNew{} of their 25 failing streaks), and their failing streaks mostly come from the project, above all from failing tests. In contrast, \nLogSameSameOtherPurposes{} of the 25 failing streaks that start and end on the same commit come from workflows that neither build nor test code, mostly (9 out of 16) workflows that handle pull requests and issues, and these mostly fail for workflow or infrastructure reasons. As in $RQ_2$, this goes together with the trigger, where \nLogSameSameOtherPurposesScheduleOrWorkflowRun{} of these \nLogSameSameOtherPurposes{} failing streaks start with a \texttt{workflow\_run} or a \texttt{schedule}.
As in the study of Zheng~\etal~\cite{zheng2025ghafail}, failing tests are the most common project cause. However, the authors found that the project itself caused 84\% of the failed runs they analysed, whereas in our sample the workflow or its infrastructure causes about as many failing streaks as the project (\nLogWorkflowAll{} against \nLogProjectAll{}).
Part of this difference comes from workflows that neither build nor test code, which mostly fail for workflow or infrastructure reasons.
It might also be due to differences in the studied populations:~we only consider runs on the default branch and include all programming languages, whereas Zheng~\etal only studied Java projects.

\begin{table}[!t]
  \caption{Cause of the first failing run of the \nLogStreaks{} sampled failing streaks, per group: the commit at the start (top) and at the end (bottom), with 25 failing streaks in each group.}
  \label{tab:log-causes}
  \centering
  \setlength{\tabcolsep}{3pt}
  \begin{tabular}{@{}lccccc@{}}
    cause & {\shortstack{new\\$\downarrow$\\new}} & {\shortstack{new\\$\downarrow$\\same}} & {\shortstack{same\\$\downarrow$\\new}} & {\shortstack{same\\$\downarrow$\\same}} & all \\
    \midrule
    \textbf{project}                     & \textbf{\nLogProjectNewNew{}} & \textbf{\nLogProjectNewSame{}} & \textbf{\nLogProjectSameNew{}} & \textbf{\nLogProjectSameSame{}} & \textbf{\nLogProjectAll{}} \\
    \quad tests fail                     & \nLogTestsFailNewNew{} & \nLogTestsFailNewSame{} & \nLogTestsFailSameNew{} & \nLogTestsFailSameSame{} & \nLogTestsFailAll{} \\
    \quad other project causes           & \nLogRestProjectNewNew{} & \nLogRestProjectNewSame{} & \nLogRestProjectSameNew{} & \nLogRestProjectSameSame{} & \nLogRestProjectAll{} \\
    \textbf{workflow / infrastructure}   & \textbf{\nLogWorkflowNewNew{}} & \textbf{\nLogWorkflowNewSame{}} & \textbf{\nLogWorkflowSameNew{}} & \textbf{\nLogWorkflowSameSame{}} & \textbf{\nLogWorkflowAll{}} \\
    \quad network issue or external service    & \nLogNetworkOrServiceNewNew{} & \nLogNetworkOrServiceNewSame{} & \nLogNetworkOrServiceSameNew{} & \nLogNetworkOrServiceSameSame{} & \nLogNetworkOrServiceAll{} \\
    \quad workflow configuration         & \nLogWorkflowConfigurationNewNew{} & \nLogWorkflowConfigurationNewSame{} & \nLogWorkflowConfigurationSameNew{} & \nLogWorkflowConfigurationSameSame{} & \nLogWorkflowConfigurationAll{} \\
    \quad other workflow causes          & \nLogRestWorkflowNewNew{} & \nLogRestWorkflowNewSame{} & \nLogRestWorkflowSameNew{} & \nLogRestWorkflowSameSame{} & \nLogRestWorkflowAll{} \\
    \textbf{other / cannot tell}         & \textbf{\nLogOtherNewNew{}} & \textbf{\nLogOtherNewSame{}} & \textbf{\nLogOtherSameNew{}} & \textbf{\nLogOtherSameSame{}} & \textbf{\nLogOtherAll{}} \\
  \end{tabular}
\end{table}

We observe that workflow or infrastructure issues are most often related to network issues or external services (\nLogNetworkOrServiceAll{} failing streaks), rather than the workflow configuration (\nLogWorkflowConfigurationAll{}). This matches the low share of commits that change workflow files at the start and end of failing streaks ($RQ_2$ and the commit analysis above).

\summary{More than a quarter of the failing streaks start and end on the same commit, without any change to the repository. In a sample of 100 failing streaks, those that end on a new commit most often fail on tests, while those that end on the same commit mostly fail because of the workflow or its infrastructure, most often a network issue or an external service.}

%% file: Input/discussion.tex
\section{Discussion}\label{sec:discussion}

We presented preliminary results to the \github product development team. They confirmed that they regularly face issues identified in our study, such as flakiness, needless reruns, and scheduled workflows that keep running while failing, but noted that no large-scale study of these problems was available. This confirmed the need for the analyses we report.
Our results have several implications for \github, for tool builders, for maintainers of workflow files, and for researchers.

Most failing streaks are short, but the \pctStreaksLongerThanDay{}\% that last more than a day make up \pctFailingTimeLongerThanDay{}\% of the time that workflows spend failing ($RQ_3$), and \pctWorkflowsAlwaysFail{}\% of workflows fail on every run ($RQ_1$). Rerunning on the same commit rarely helps after it has failed several times. After a first failure in a failing streak, the next run on the same commit passes in \pctSameCommitPassesAfterOne{}\% of cases; after six or more failures, it passes in only \pctSameCommitPassesAfterSixPlus{}\% of cases.
One might expect long failing streaks to belong to abandoned projects, but \pctNewCommitLongSchedule{}\% of the scheduled failing streaks that last more than a week received new commits while they were still failing ($RQ_3$).
However, \github notifies the failures of a scheduled workflow only to the user who last changed its cron schedule~\cite{GitHubDocsNotifications}, who may no longer be active in the project.
\summary{
\github could therefore warn the active developers of a repository when a workflow is still failing after a day, and pause the scheduled runs that keep failing on the same commit until the code or the workflow changes. Alternatively, maintainers can expose such failures themselves, for example with a step that opens an issue when a scheduled run fails.
}

More than one in three failing streaks start on the same commit as the last successful run ($RQ_2$), and two in five end on the same commit as the last failing run ($RQ_4$). In these cases, the same commit both failed and passed, so repository changes alone do not explain the failure.
In our log sample, failing streaks that end on the same commit mostly fail due to the workflow or its infrastructure, often due to a network issue or an external service ($RQ_{4}$).
Tests can also fail and then pass on unchanged code ($RQ_4$), suggesting flaky tests~\cite{ge2026understandingdetectingflakybuilds}.
\summary{Maintainers could check whether the same commit has passed before and rerun the failed jobs. Tool builders could identify flaky failures from their logs and suggest reruns rather than fixes. \github could retry these failing runs automatically or show what changed since the last successful run beyond the code, such as the runner image or action versions. Runs of the same commit that both fail and pass provide tool builders and researchers with many examples of flaky failures.}

Most commits at the start and end of failing streaks change project files only. Still, the file defining the workflow is changed in \pctBreakPairsOwnWorkflowFile{}\% of the commits that start a failing streak and in \pctFixPairsOwnWorkflowFile{}\% of those that end one, compared with \pctBaselinePairsOwnWorkflowFile{}\% of the changes between two successful runs ($RQ_2$, $RQ_4$).
Changing a workflow is thus more common at the start and end of failing streaks.

\summary{Maintainers could check workflow changes with a linter such as actionlint~\cite{actionlint} before merging them, or run the changed workflow on a branch first. \github could facilitate this process, \eg by letting maintainers test a changed workflow before it reaches the default branch.}

Finally, our results have implications for researchers.
Failing runs mostly occur in streaks, since \pctFailingRunsAfterFailingRun{}\% of them follow another failing run of the same workflow ($RQ_1$), and a failing streak usually keeps the same cause until it ends ($RQ_{4}$). A random sample of failing runs therefore mostly contains repeated failures, making the failing streak a more natural unit for studying workflow failures.
Workflow purpose also matters: workflows that build or test code mostly fail because of the project, while workflows with other purposes mostly fail because of the workflow or its infrastructure, partly explaining why our findings on causes differ from those of Zheng~\etal~\cite{zheng2025ghafail}.
Multi-purpose workflows also have failing streaks more often, even at the same size ($RQ_1$); whether splitting them into single-purpose workflows would reduce failing streaks remains an open question. In addition, because the \github API only lists the last attempt of a run, we retrieved \nHiddenAttempts{} earlier attempts. Without them, we would have missed failing streaks ending with a rerun.
\summary{Empirical studies of workflow failures should analyse failing streaks in complement to individual failing runs. They should account for workflow purpose, and include rerun attempts.
}

%% file: Input/threats.tex
\section{Threats to Validity}
\label{sec:threats}

We discuss the threats to validity of our research following the structure recommended by Wohlin \etal~\cite{wohlin2012}.

\textbf{Construct validity.} %
We keep only runs that succeeded or failed. After applying filters (1) to (3) in \sect{sec:dataExtraction}, \pctCompletedRunsSkipped{}\% of the completed runs had been skipped and \pctCompletedRunsActionRequired{}\% were waiting for approval; these runs did not execute, so removing them does not change the failing streaks. Another \pctCompletedRunsCancelled{}\% were cancelled before completion; some of them may hide a failure, so we may miss some failing runs.

The LLM may assign wrong purposes to some workflows. Its agreement with our consolidated labels of 100 workflows is $\alpha=0.76$, and for 18 of the 32 workflows where we disagreed, its purpose set was the correct one (\sect{sec:classification}). We also labelled each workflow from the latest version of its file, whose purposes may differ from those it had during a failing streak.

\textbf{Internal validity.} %
We excluded the \pctFailingStreaksDropped{}\% of failing streaks that were already underway when data collection started or still underway when it ended, because we cannot observe their start or end. Long failing streaks are more likely to be excluded, which matters most for durations ($RQ_3$). We therefore also counted the excluded failing streaks over the part of the window that we observe: the failing streaks longer than a day then hold \pctFailingTimeLongerThanDayWithDropped{}\% of the failing time instead of \pctFailingTimeLongerThanDay{}\%, so excluding them makes this result an underestimate.

Whether a failing streak starts or ends on a new commit does not show that this commit caused or fixed the failure: a new commit may simply come with the next run without fixing anything, and a failure on an unchanged commit may still come from an earlier change, such as a flaky test.

Relying on an LLM is another choice that may affect our results. Its run-to-run variation does not depend on the failing streaks, since the LLM only sees the workflow file, so it would tend to blur rather than create the differences between purposes that we report; the effect of our prompt is assessed by the comparison with our manual labels (\sect{sec:classification}).

\textbf{Conclusion validity.} %
Our results on failing streaks count each failing streak once, so workflows with many failing streaks weigh more. With one random failing streak per workflow, the median duration rises from \medianFailingHours{} to \medianFailingHoursPerWorkflow{} hours, which is why $RQ_3$ reports both views.

Apart from the log analysis of $RQ_4$ (100 failing streaks, 25 per group), our results cover all runs in scope. Its pooled counts over-represent the failing streaks that start or end on the same commit, which make up three quarters of the sample but about half of all failing streaks. Weighting each group by its share of all failing streaks gives about the same picture as the raw counts (\nLogProjectAll{} and \nLogWorkflowAll{} of 100): \pctLogProjectWeighted{}\% project causes and \pctLogWorkflowWeighted{}\% workflow or infrastructure causes.
Each failing streak was labelled by one author, but both authors followed the same guide, used the failure categories of Zheng~\etal~\cite{zheng2025ghafail}, and compared each failing log with the log of the successful run before it.

\textbf{External validity.} %
The GHARuns dataset only covers public \github repositories with at least 100 stars and 300 commits~\cite{GHARuns2026}, and we applied further filters (\sect{sec:dataExtraction}). Our findings may therefore not generalise to other repositories (\eg private, less popular, or less active ones), other runs (\eg pull-request runs), or periods other than our \windowDaysShort{} days.

%% file: Input/conclusion.tex
\section{Conclusion}\label{sec:conclusion}

We studied failing streaks in \nRunsInScopeShort{} runs of \nWorkflowsInScopeShort{} \gha workflows in \nReposInScopeShort{} repositories. Three in four failing runs follow another failing run of the same workflow, and the one in five failing streaks that last more than a day account for over 90\% of the time that workflows spend failing. A third of failing streaks start, and two in five end, without any change to the repository. Multi-purpose workflows have failing streaks more often.

In future work, we plan to classify failure causes automatically from the jobs, steps, and logs of the runs, to reveal at scale why workflows fail. Linking failing streaks to the contents and changes of workflow files would show which parts of a workflow make failures more likely, and whether splitting a multi-purpose workflow into single-purpose ones reduces failing streaks. Finally, we plan to survey the maintainers of workflows with long failing streaks, to learn whether they knew that the workflow was failing, what caused the failure in their view, and whether a warning after a day would have helped.

%% file: Input/data_availability.tex
\section{Data Availability}\label{sec:data_availability}
Our replication package~\cite{replicationPackage} contains all data and code needed to replicate our results, including the LLM prompt and its output, our manual labels, and the log analysis data.

%% file: msr2027.bbl
\begin{thebibliography}{10}
\providecommand{\url}[1]{#1}
\csname url@samestyle\endcsname
\providecommand{\newblock}{\relax}
\providecommand{\bibinfo}[2]{#2}
\providecommand{\BIBentrySTDinterwordspacing}{\spaceskip=0pt\relax}
\providecommand{\BIBentryALTinterwordstretchfactor}{4}
\providecommand{\BIBentryALTinterwordspacing}{\spaceskip=\fontdimen2\font plus
\BIBentryALTinterwordstretchfactor\fontdimen3\font minus
  \fontdimen4\font\relax}
\providecommand{\BIBforeignlanguage}[2]{{%
\expandafter\ifx\csname l@#1\endcsname\relax
\typeout{** WARNING: IEEEtran.bst: No hyphenation pattern has been}%
\typeout{** loaded for the language `#1'. Using the pattern for}%
\typeout{** the default language instead.}%
\else
\language=\csname l@#1\endcsname
\fi
#2}}
\providecommand{\BIBdecl}{\relax}
\BIBdecl

\bibitem{Zampetti2021}
F.~Zampetti, S.~Geremia, G.~Bavota, and M.~Di~Penta, ``{CI/CD} pipelines
  evolution and restructuring: A qualitative and quantitative study,'' in
  \emph{Int'l Conf. Software Maintenance and Evolution ({ICSME})}, 2021.

\bibitem{Golzadeh2022SANER}
M.~Golzadeh, A.~Decan, and T.~Mens, ``On the rise and fall of {CI} services in
  {GitHub},'' in \emph{Int'l Conf. Software Analysis, Evolution and
  Reengineering}.\hskip 1em plus 0.5em minus 0.4em\relax IEEE, 2022.

\bibitem{zheng2025ghafail}
L.~Zheng, S.~Li, X.~Huang, J.~Huang, B.~Lin, J.~Chen, and J.~Xuan, ``Why do
  {GitHub Actions} workflows fail? {An} empirical study,'' \emph{ACM Trans.
  Softw. Eng. Methodol.}, Jul. 2025.

\bibitem{GHARuns2026-dataset}
\BIBentryALTinterwordspacing
A.~Talebzadeh~Bardsiri, T.~Mens, and A.~Decan, ``{GHARuns}: A large dataset of
  github actions workflow execution metadata,'' in \emph{Int'l Conf. Software
  Maintenance and Evolution ({ICSME})}.\hskip 1em plus 0.5em minus 0.4em\relax
  IEEE, 2026. [Online]. Available:
  \url{https://orbi.umons.ac.be/handle/20.500.12907/56424}
\BIBentrySTDinterwordspacing

\bibitem{hilton2016usage}
M.~Hilton, T.~Tunnell, K.~Huang, D.~Marinov, and D.~Dig, ``Usage, costs, and
  benefits of continuous integration in open-source projects,'' in \emph{Int'l
  Conf. Automated Software Engineering ({ASE})}.\hskip 1em plus 0.5em minus
  0.4em\relax IEEE, 2016, pp. 426--437.

\bibitem{Zampetti2020badpractices}
F.~Zampetti, C.~Vassallo, S.~Panichella, G.~Canfora, H.~Gall, and M.~Di~Penta,
  ``An empirical characterization of bad practices in continuous integration,''
  \emph{Empirical Software Engineering}, vol.~25, 2020.

\bibitem{Huang2026GHAReruns}
J.~Huang and B.~Lin, ``On the reruns of {GitHub Actions} workflows,'' \emph{ACM
  Trans. Softw. Eng. Methodol.}, Feb. 2026.

\bibitem{ge2026understandingdetectingflakybuilds}
\BIBentryALTinterwordspacing
W.~Ge and C.~Zhang, ``Understanding and detecting flaky builds in {GitHub
  Actions},'' 2026, arXiv preprint. [Online]. Available:
  \url{https://doi.org/10.48550/arXiv.2602.02307}
\BIBentrySTDinterwordspacing

\bibitem{Talebzadeh2026-ICSME}
A.~Talebzadeh~Bardsiri, A.~Decan, and T.~Mens, ``On the {GitHub Actions}
  language: Usage, evolution, and workflow reliability,'' in \emph{Int'l Conf.
  Software Maintenance and Evolution}.\hskip 1em plus 0.5em minus 0.4em\relax
  IEEE, 2026.

\bibitem{Zhao2017-CI}
Y.~Zhao, A.~Serebrenik, Y.~Zhou, V.~Filkov, and B.~Vasilescu, ``The impact of
  continuous integration on other software development practices: A large-scale
  empirical study,'' in \emph{Int'l Conf. Automated Software Engineering
  ({ASE})}, 2017, pp. 60--71.

\bibitem{Rostami2023CI}
P.~Rostami~Mazrae, T.~Mens, M.~Golzadeh, and A.~Decan, ``On the usage, co-usage
  and migration of {CI/CD} tools: A qualitative analysis,'' \emph{Empir. Softw.
  Eng.}, vol.~28, no.~2, p.~52, 2023.

\bibitem{Onsori2026}
H.~Onsori~Delicheh, G.~Cardoen, A.~Decan, and T.~Mens, ``Automation and reuse
  practices in {GitHub Actions} workflows: A practitioner's perspective,''
  \emph{ACM Trans. Softw. Eng. Methodol.}, 2026.

\bibitem{Kinsman2021}
T.~Kinsman, M.~Wessel, M.~A. Gerosa, and C.~Treude, ``How do software
  developers use {GitHub Actions} to automate their workflows?'' in \emph{Int'l
  Conf. Mining Software Repositories}, 2021.

\bibitem{decan2022gha}
A.~Decan, T.~Mens, P.~Rostami~Mazrae, and M.~Golzadeh, ``On the use of {GitHub
  Actions} in software development repositories,'' in \emph{Int'l Conf.
  Software Maintenance and Evolution}.\hskip 1em plus 0.5em minus 0.4em\relax
  IEEE, 2022.

\bibitem{DECAN2023111827}
A.~Decan, T.~Mens, and H.~Onsori~Delicheh, ``On the outdatedness of workflows
  in the {GitHub Actions} ecosystem,'' \emph{J. Systems and Software}, vol.
  206, 2023.

\bibitem{Khatami2024CatchingSmells}
A.~Khatami, C.~Willekens, and A.~Zaidman, ``Catching smells in the act: A
  {GitHub Actions} workflow investigation,'' in \emph{Int'l Conf. Source Code
  Analysis and Manipulation}.\hskip 1em plus 0.5em minus 0.4em\relax IEEE,
  2024.

\bibitem{ghaleb2025complexity}
T.~Ghaleb, O.~Abduljalil, and S.~Hassan, ``{CI/CD} configuration practices in
  open-source {Android} apps: An empirical study,'' \emph{ACM Trans. Softw.
  Eng. Methodol.}, May 2025.

\bibitem{rostami2026GHAEvolution}
P.~Rostami~Mazrae, A.~Decan, T.~Mens, and M.~Wessel, ``An empirical study of
  the evolution of {GitHub Actions} workflows,'' \emph{Journal of Systems and
  Software}, vol. 236, 2026.

\bibitem{Beller2017}
M.~Beller, G.~Gousios, and A.~Zaidman, ``Oops, my tests broke the build: An
  explorative analysis of {Travis CI} with {GitHub},'' in \emph{Int'l Conf.
  Mining Software Repositories}, 2017, pp. 356--367.

\bibitem{Rausch2017BuildFailures}
T.~Rausch, W.~Hummer, P.~Leitner, and S.~Schulte, ``An empirical analysis of
  build failures in the continuous integration workflows of {Java}-based
  open-source software,'' in \emph{International Conference on Mining Software
  Repositories ({MSR})}, 2017, pp. 345--355.

\bibitem{Gallaba2018NoiseHeterogeneity}
K.~Gallaba, C.~Macho, M.~Pinzger, and S.~McIntosh, ``Noise and heterogeneity in
  historical build data: An empirical study of {Travis CI},'' in
  \emph{International Conference on Automated Software Engineering ({ASE})},
  2018, pp. 87--97.

\bibitem{Jin2020CostEfficientCI}
\BIBentryALTinterwordspacing
X.~Jin and F.~Servant, ``A cost-efficient approach to building in continuous
  integration,'' in \emph{International Conference on Software Engineering
  ({ICSE})}.\hskip 1em plus 0.5em minus 0.4em\relax ACM, 2020, pp. 13--25.
  [Online]. Available: \url{https://doi.org/10.1145/3377811.3380437}
\BIBentrySTDinterwordspacing

\bibitem{shirakawa2026TestAlertSnooze}
\BIBentryALTinterwordspacing
A.~Shirakawa, T.~Shirai, Y.~Kashiwa, M.~Kondo, Y.~Kamei, and H.~Iida, ``Test
  alert snooze: An empirical study of consecutive test failures on {CI},'' in
  \emph{International Symposium on Empirical Software Engineering and
  Measurement ({ESEM})}, ser. Leibniz International Proceedings in Informatics
  (LIPIcs), vol. 394, 2026. [Online]. Available:
  \url{https://drops.dagstuhl.de/entities/document/10.4230/LIPIcs.ESEM.2026.24}
\BIBentrySTDinterwordspacing

\bibitem{Khatami2026EASE}
A.~Khatami, C.~Brandt, and A.~Zaidman, ``Beyond the {YAML} file: Understanding
  real-world {GitHub Actions} workflow adoption,'' in \emph{Int'l Conf.
  Evaluation and Assessment in Software Engineering}.\hskip 1em plus 0.5em
  minus 0.4em\relax IEEE, 2026.

\bibitem{GitHubDocsEvents}
{GitHub}, ``Events that trigger workflows,''
  \url{https://docs.github.com/en/actions/reference/workflows-and-actions/events-that-trigger-workflows},
  accessed: 2026-10-23.

\bibitem{GHARuns2026}
\BIBentryALTinterwordspacing
A.~Talebzadeh~Bardsiri, T.~Mens, and A.~Decan, ``Gharuns: A large dataset of
  github actions workflow execution metadata,'' Oct. 2026. [Online]. Available:
  \url{https://doi.org/10.5281/zenodo.23458821}
\BIBentrySTDinterwordspacing

\bibitem{GitHubDocsRerun}
{GitHub}, ``Re-run workflows and jobs,''
  \url{https://docs.github.com/en/actions/how-tos/manage-workflow-runs/re-run-workflows-and-jobs},
  accessed: 2026-10-21.

\bibitem{hayes2007answering}
A.~F. Hayes and K.~Krippendorff, ``Answering the call for a standard
  reliability measure for coding data,'' \emph{Communication methods and
  measures}, vol.~1, no.~1, pp. 77--89, 2007.

\bibitem{GitHubDocsNotifications}
{GitHub}, ``Notifications for workflow runs,''
  \url{https://docs.github.com/en/actions/concepts/workflows-and-actions/notifications-for-workflow-runs},
  accessed: 2026-10-17.

\bibitem{GitHubDocsManageNotifications}
------, ``Managing {GitHub Actions} notifications,''
  \url{https://docs.github.com/en/subscriptions-and-notifications/how-tos/managing-github-actions-notifications},
  accessed: 2026-10-18.

\bibitem{GitHubDocsRetention}
------, ``Configuring the retention period for artifacts, and logs,''
  \url{https://docs.github.com/en/organizations/managing-organization-settings/configuring-the-retention-period-for-github-actions-artifacts-and-logs-in-your-organization},
  accessed: 2026-10-21.

\bibitem{Hubner2025CiDiff}
N.~Hubner, J.-R. Falleri, R.~Uricaru, T.~Degueule, and T.~Durieux, ``What
  happened in this pipeline? {Diffing} build logs with {CiDiff},'' \emph{Proc.
  ACM Softw. Eng.}, vol.~2, no. ISSTA089, Jun. 2025.

\bibitem{actionlint}
{rhysd}, ``actionlint: Static checker for {GitHub Actions} workflow files,''
  \url{https://github.com/rhysd/actionlint}, accessed: 2026-10-23.

\bibitem{wohlin2012}
C.~Wohlin, P.~Runeson, M.~H\"ost, M.~C. Ohlsson, B.~Regnell, and A.~Wessl\'en,
  \emph{Experimentation in Software Engineering}.\hskip 1em plus 0.5em minus
  0.4em\relax Springer, 2012.

\bibitem{replicationPackage}
Anonymous, ``Replication package,''
  \url{https://figshare.com/s/fb59be9661cb01975555}, 2026.

\end{thebibliography}
